\documentclass[runningheads]{llncs}
\usepackage[T1]{fontenc}
\usepackage{graphicx}
\usepackage{booktabs}
\usepackage{multirow}
\usepackage{subfigure}
\usepackage{comment}
\begin{document}
\title{Perceptual Analysis of Groups of Virtual Humans Animated using Interactive Platforms}
\titlerunning{Perceptual Analysis of Groups of Virtual Humans}
%
\author{Rubens Montanha \and
Giovana Raupp \and 
Ana Carolina Schmitt \and 
Gabriel Schneider \and
Victor Araujo \and
Soraia Raupp Musse\orcidID{0000-0002-3278-217X}}
\authorrunning{R. Montanha et al.}
%
\institute{VHLab, Graduate Course in Computer Science \\
Pontifical Catholic University of Rio Grande do Sul, Brazil \\
\email{soraia.musse@pucrs.br}\\
\url{www.inf.pucrs.br/vhlab} }
%
\maketitle              
\begin{abstract}
Virtual humans (VH) have been used in Computer Graphics (CG) for many years, and perception studies have been applied to understand how people perceive them. Some studies have already examined how realism impacts the comfort of viewers. In some cases, the user's comfort is related to human identification. For example, people from a specific group may look positively at others from the same group. Gender is one of those characteristics that have in-group advantages. For example, in terms of VHs, studies have shown that female humans are more likely to recognize emotions in female VHs than in male VHs. However, there are many other variables that can impact the user perception. To aid this discussion, we conducted a study on how people perceive comfort and realism in relation to interactive VHs with different genders and expressing negative, neutral, or positive emotions in groups. We created a virtual environment for participants to interact with groups of VHs, which are interactive and should evolve in real-time, using a popular game engine. To animate the characters, we opted for cartoon figures that are animated by tracking the facial expressions of actors, using available game engine platforms to conduct the driven animation. Our results indicate that the emotion of the VH group impacts both comfort and realism perception, even by using simple cartoon characters in an interactive environment. Furthermore, the findings suggest that individuals reported feeling better with a positive emotion compared to a negative emotion, and that negative emotion recognition is impacted by the gender of the VHs group. Additionally, although we used simple characters, the results are consistent with the perception obtained when analysing realistic the state-of-the-art virtual humans, which positive emotions tend to be more correctly recognized than negative ones.

\end{abstract}
\section{Introduction}
\label{sec:introduction}
The use of Computer Graphics (CG) has become increasingly popular in recent years, particularly in games and movies. In many of such applications, VHs utilize body and face capture techniques to enhance the realism of characters' behaviors. With these advancements, Virtual Humans (VH) can convey comfort, persuasion, and affection to individuals who watch or interact with them
~\cite{katsyri2015review}. However, some people may also feel negative emotional valences, as discussed in Mori's work~\cite{mori2012uncanny} known as the Uncanny Valley (UV). According to Mori, the UV is a phenomenon where a human response may fall into a valley of discomfort/uncanny depending on how much the VH resembles a real human.
While Mori's theory was initially applied to robots, various authors have explored the impact of the UV on CG characters~\cite{araujo2021perceived,draude2011intermediaries,macdorman2016reducing}. 
According to Katsyri et al.~\cite{katsyri2015review}, the feeling of discomfort is associated with human identification, wherein individuals observing or interacting with VH seek human characteristics in them. One such characteristic is gender. For example, in a recent study by Araujo et al.~\cite{araujo2021analysis}, 
the authors noted that women reported feeling more comfortable with very realistic female characters than with very realistic male characters. Conversely, men expressed a similar level of comfort with both gender characters. 
The gender of the VH may influence the perception of different attributes, e.g., in the studies of Zibrek et al.~\cite{zibrek2013evaluating,zibrek2015exploring}, the results showed that appearance, walking movement, and speeches of the VHs, animated by real actors and actresses, can influence the emotional perception of gender.

Just like gender, facial emotions also represent an important human characteristic for non-verbal communication, and have been extensively studied in the literature of VHs~\cite{zibrek2013evaluating,queiroz2014investigating,andreotti2021perception}. Among the techniques available for facial animation, we can mention the use of blend shapes~\cite{joshi2006learning}, which enables the linear combination of facial movements. Blend shapes have been employed in conjunction with the Facial Action Coding System (FACS)~\cite{alkawaz2015blend}, a system proposed by Ekman and Friesen~\cite{ekman1978facial}, that specifies and parameterizes different parts of the human face. 
In the work of Tinwell et al.~\cite{tinwell2011facial}, the authors measured the perceived uncanny feeling in relation to all Ekman emotions expressed by VHs. The results showed that negative emotions tended to convey feelings of strangeness to people, except for the emotion of anger. While happiness did not cause feelings of strangeness.

In the work of Zojaji and colleagues~\cite{zojaji2020influence}, the use of verbal and nonverbal behavior 
by VHs can lead to different user experiences. For example, the authors stated that when a newcomer was invited to join a group of virtual agents, employing direct and explicit politeness strategies proved to be more persuasive.
Kendon~\cite{kendon1990conducting} described a common space management concept in a group, which is divided into three spaces: o-space, p-space, and r-space. During an interaction, the group of people forms a convex empty area known as the o-space, surrounded by the individuals involved in the interaction, while the p-space is where the group participants are located. The r-space extends beyond the p-space. 
In a group, crowd perception allows the observer to perceive people's behavior and communication at a collective level~\cite{elias2017ensemble}. In that case, the observer can discern distinct traces and evaluate patterns among various groups within a crowd~\cite{campbell1958common}. Certain emergent properties can be observed in groups but not in individuals, including patterns of coordinated behavior and expressions~\cite{lamer2018rapid}.

In this study, we aim to investigate the impact of simple cartoons and interactive VHs on human perception. Specifically, we want to understand how the gender and emotion of interactive and simple cartoons characters in a group can affect people's perceptions of comfort and realism.
To answer this question, this work introduces a study on how people feel when interacting with a group of cartoons VHs with different genders expressing positive, neutral, or negative emotions animated using real time platforms to extract data from video. Accordingly, three hypotheses were formulated:
\begin{itemize}
    \item $H0_1$ - People recognize emotions similarly in groups of VHs with different genders and emotions.
    \item $H0_2$ - People experience similar comfort towards groups of VHs with different genders and emotions.
    \item $H0_3$ - People perceive the realism of groups of VHs with different genders and emotions similarly.
\end{itemize}

To investigate people's perception of interactive VHs, we designed an experiment in an interactive environment to address the hypotheses related to gender and emotions.


\section{Related Work}
\label{sec:related_work}

This section presents studies on human perception about groups~\cite{Favarettl2019}, gender, emotions, and interactive VHs in the virtual world~\cite{Tharindu2019}. According to Johansson~\cite{johansson1973visual}, humans can generally recognize/categorize human behavioral movements from little information. Emotions are also part of these human behaviors, as presented in several studies (\cite{friesen1983emfacs,EkmanFACS,Ekman2013AnAF,bassili1978facial,ekman1978facial}). Human perception has been part of several studies in the CG community, and emotional perception is also a topic related to VHs (\cite{ennis2013emotion,andreotti2021perception,queiroz2014investigating,tinwell2013perception}). Regarding gender, literature studies show that the different genders of participants can influence the recognition of emotions (\cite{zibrek2013evaluating,zibrek2015exploring,durupinar2022facial,bailey2017gender}) and the gender perception/attribution of VHs (\cite{mcdonnell2007virtual,araujo2021analysis,ghosh4evaluating}). For example, in the work of Durupinar and Kim~\cite{durupinar2022facial}, the results showed that emotions expressed by female VHs were recognized more easily than when compared to male VHs.

Based on the research conducted by Brown~\cite{brown2020social}, Social Identity theory explains that individuals tend to view themselves and their groups in a positive light. As a result, people from the same group may have certain advantages, known as in-group advantages, when it comes to categorizing, assigning, and recognizing characteristics of others who belong to the same group (\cite{guadagno2007virtual,elfenbein2002there,krumhuber2015real}). In terms of emotions and gender, for example, the work of McDuff et al.~\cite{mcduff2017large} showed that women, in comparison with men, tend to be more expressive, positively and negatively. In a study by Abbruzzese et al.~\cite{abbruzzese2019age}, it was found that women may have an advantage in recognizing emotions expressed by other women, known as in-group advantages. When it comes to groups that are different from our own, we can sometimes fall into what's known as the Dehumanization Theory (\cite{brown2020social}). This means that we may start to view people who do not look like us or are not part of our group as less human. In the context of VHs, this dehumanization could result in a reduction of the level of human-like qualities attributed to VHs that are not part of our specific group, i.e., be considered less realistic.

Then, studying group perception is essential to learning about social patterns~\cite{musse2021history,lamer2018rapid}. The area of group perception has grown in recent years in several scientific researches in Psychology and Computer Science, for example, in the work of Lamer et al.~\cite{lamer2018rapid}, the authors measured whether racial categories were influenced by faces with different emotions.
Regarding VHs, McDonnel et al.~\cite{mcdonnell2008clone} conducted research on the perception of different clones in crowd simulations. The findings showed that participants had shorter reaction times in detecting pairs of VHs when the number of clones in crowds of VHs was greater. In another study, Araujo et al.~\cite{araujo2021much} examined the human perception of geometric characteristics, personalities, and emotions in VHs. The results indicated that women perceived more positive emotions and personality traits, while men perceived more negative characteristics.

In terms of interaction with groups of VHs, in the work of Volonte et al.~\cite{volonte2020effects}, the authors examined the perception of people when interacting with a crowd of VHs in a virtual reality environment. They discovered that the participants could interpret correctly verbal and non-verbal behaviors in a crowd of VHs.

These aforementioned studies helped us model stimuli involving groups of VHs with different gender and emotion models. The purpose of these stimuli is to try to test our hypotheses. The next section presents the methodology for creating the stimuli.

\section{Methodology}
\label{sec:methodology}

To test hypotheses $H0_1$ (People recognize emotions similarly in groups of VHs with different genders and emotions), $H0_2$ (People experience similar comfort towards groups of VHs with different genders and emotions), and $H0_3$ (People perceive the realism of groups of VHs with different genders and emotions similarly), we created a virtual environment with groups of VHs, altering both the gender and emotions of the virtual characters. This section is divided into four parts: Section~\ref{sec:stimuli} explains how the stimuli were created, detailing the animation techniques, the construction of the virtual environment, including the technology used and the experiment setup. Section~\ref{sec:participants_action} presents the participants' possible actions.
Finally, Section~\ref{sec:questionnaire} explains how the questionnaire was conducted.

\begin{figure*}[!htb]
  \centering
  \subfigure[Top view of the virtual bar.]{
    \includegraphics[width=0.47\textwidth]{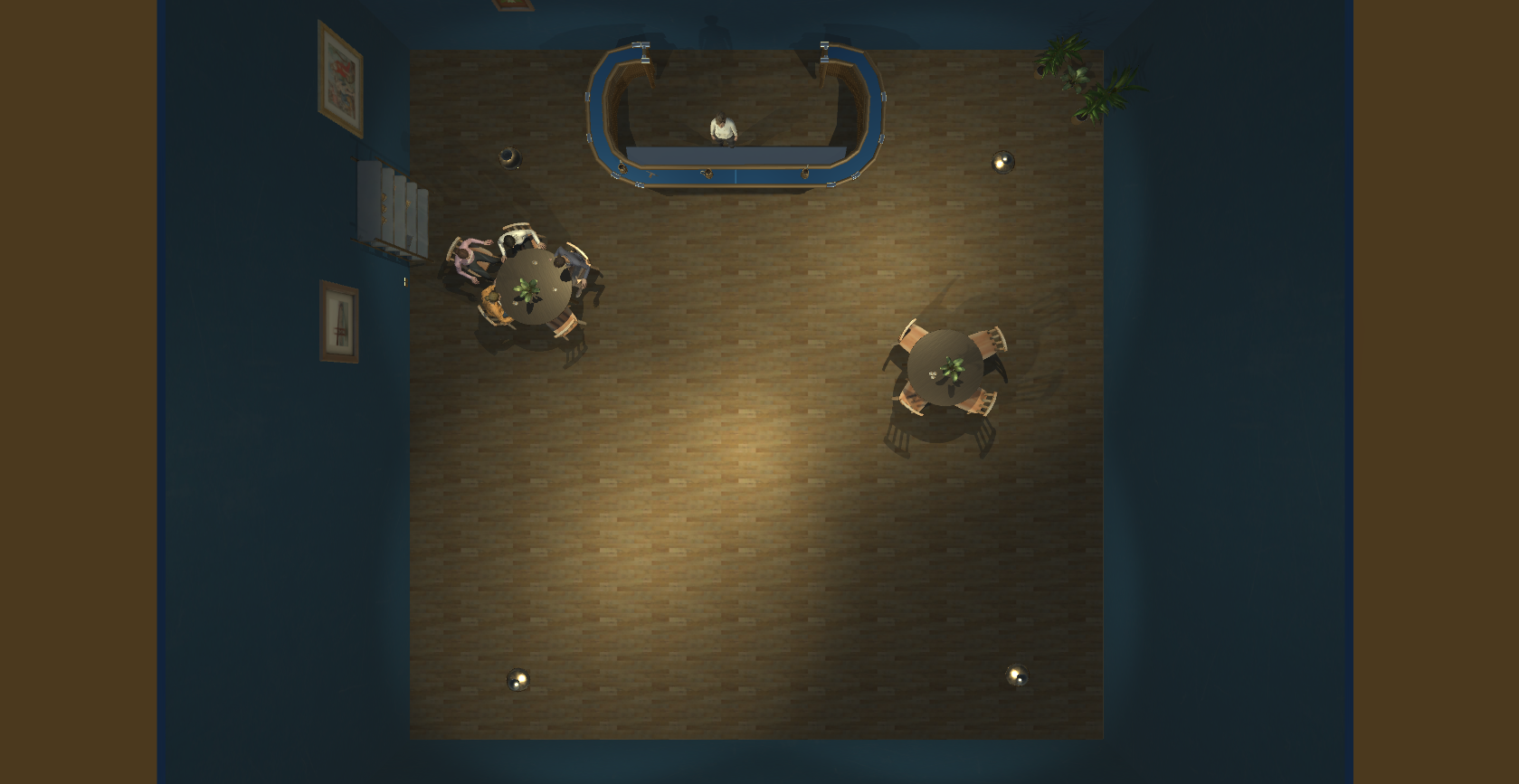}
    \label{fig:vhbar-top}
  }
  \subfigure[Corner view of the virtual bar.]{
    \includegraphics[width=0.47\textwidth]{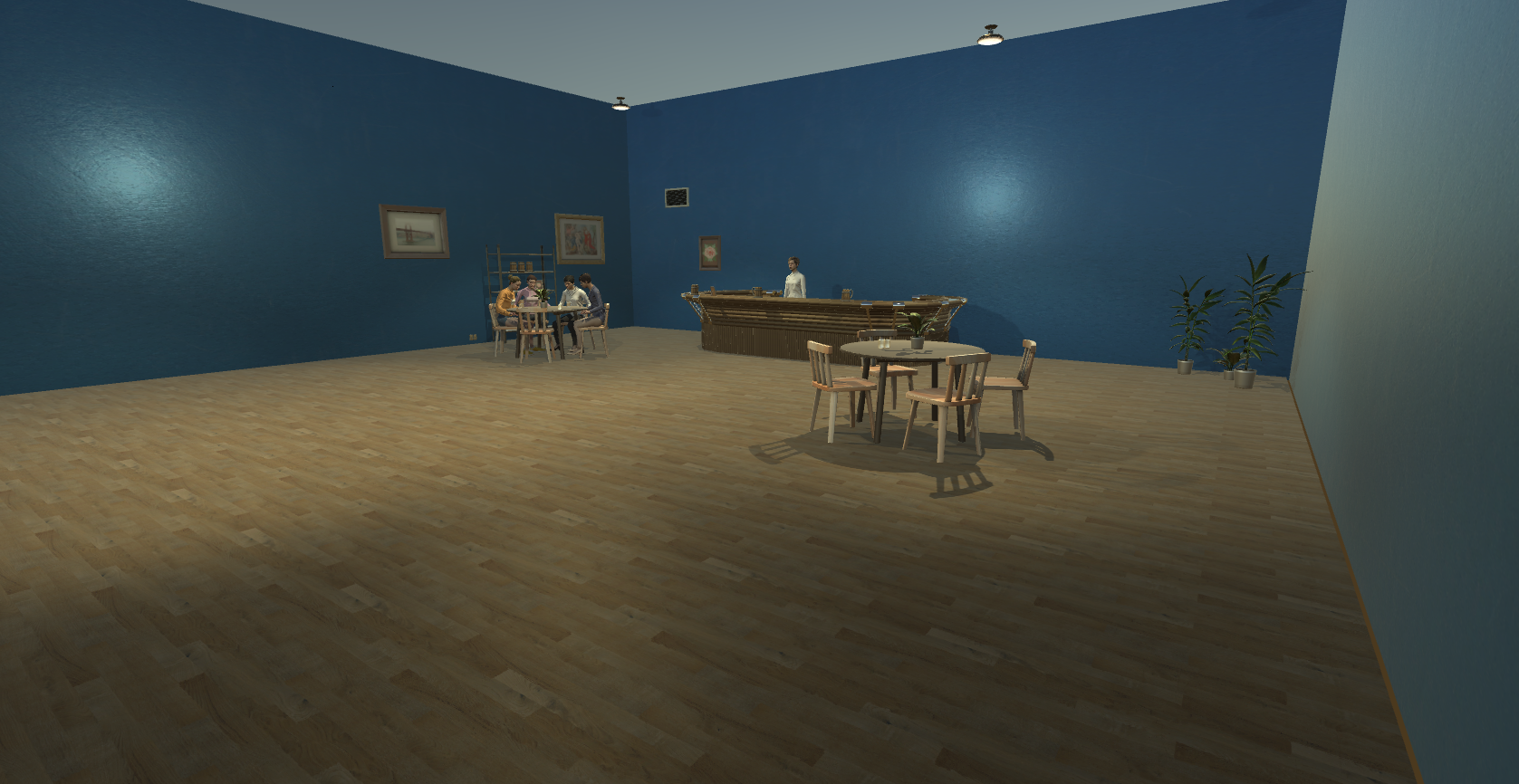}
    \label{fig:inferior-right}
  }
  \subfigure[Initial view of the virtual bar.]{
    \includegraphics[width=0.47\textwidth]{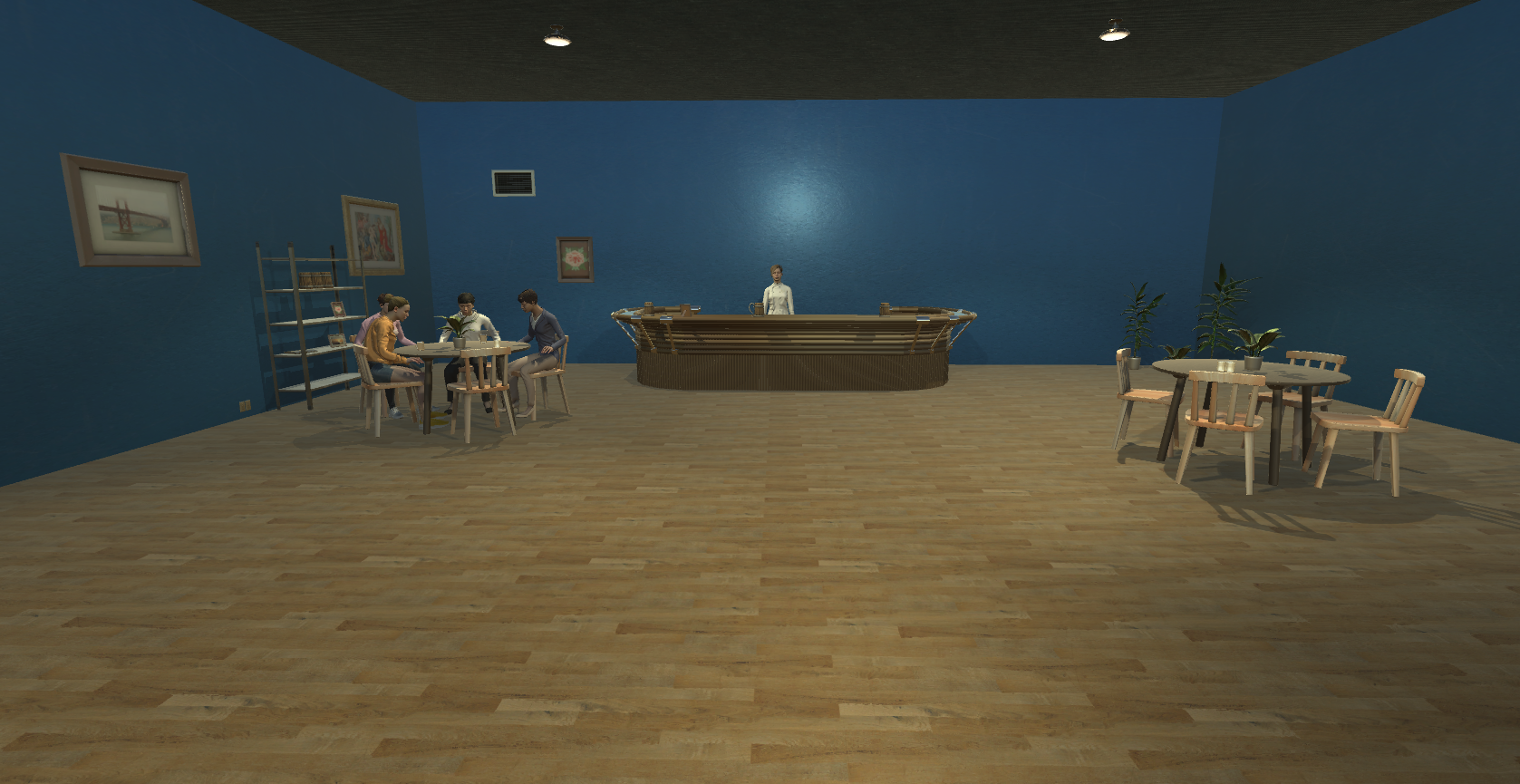}
    \label{fig:vhbar-start}
  }
  \caption{The figures present the virtual bar. Figure~\ref{fig:vhbar-top} presents the top view of the bar, showing the tables with the VHs, and the deck table with the mixologist. Figure~\ref{fig:inferior-right} presents another view from the virtual bar, while Figure~\ref{fig:vhbar-start} presents the place where the participant starts in the interaction.}
  \label{fig:vhbar}
\end{figure*}

\begin{figure*}[!htb]
    \centering
    \subfigure[Male agents at the table.]{
        \includegraphics[width=0.47\textwidth]{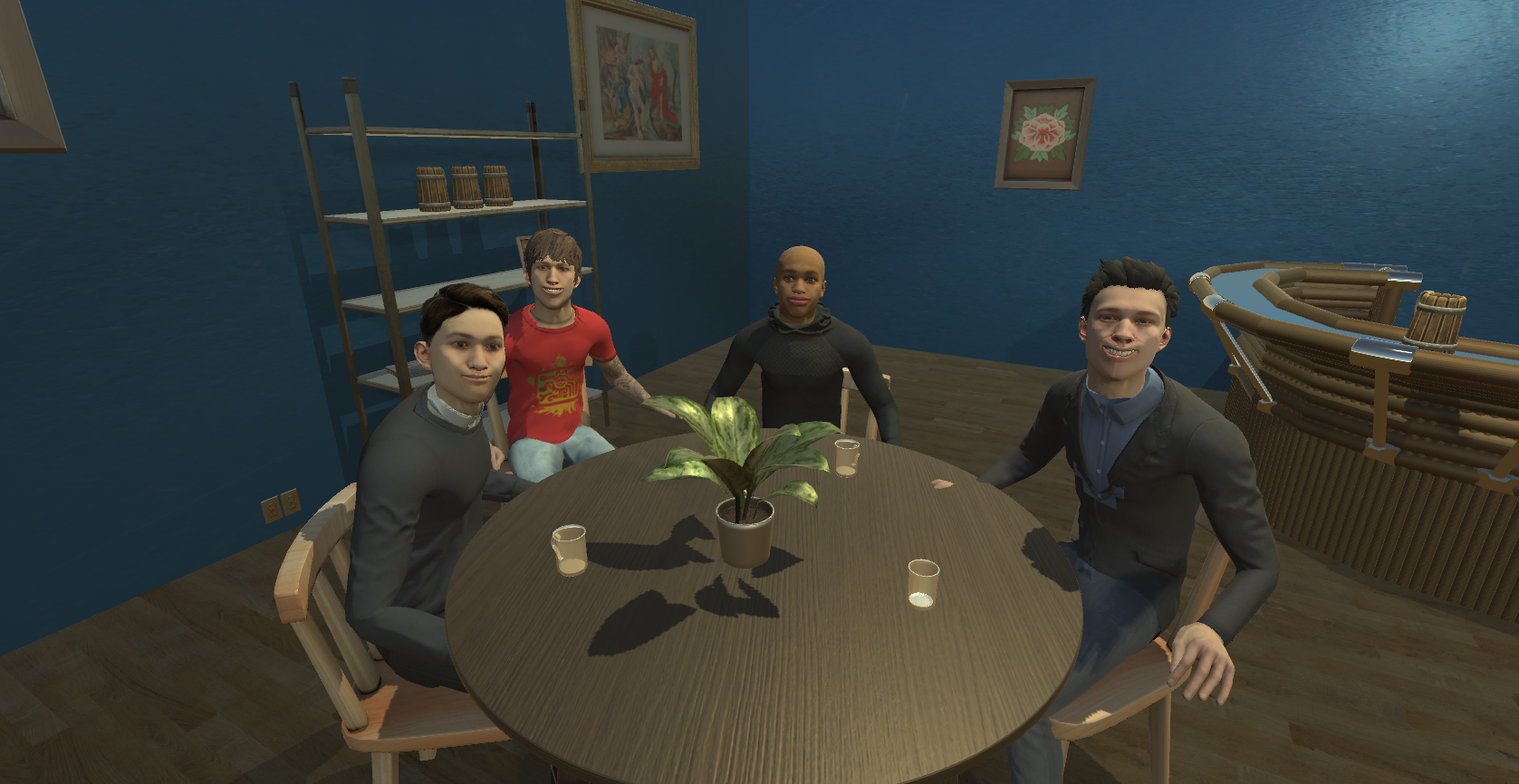}
        \label{fig:males-vhbar}
    }
    \subfigure[Female agents at the table.]{
        \includegraphics[width=0.47\textwidth]{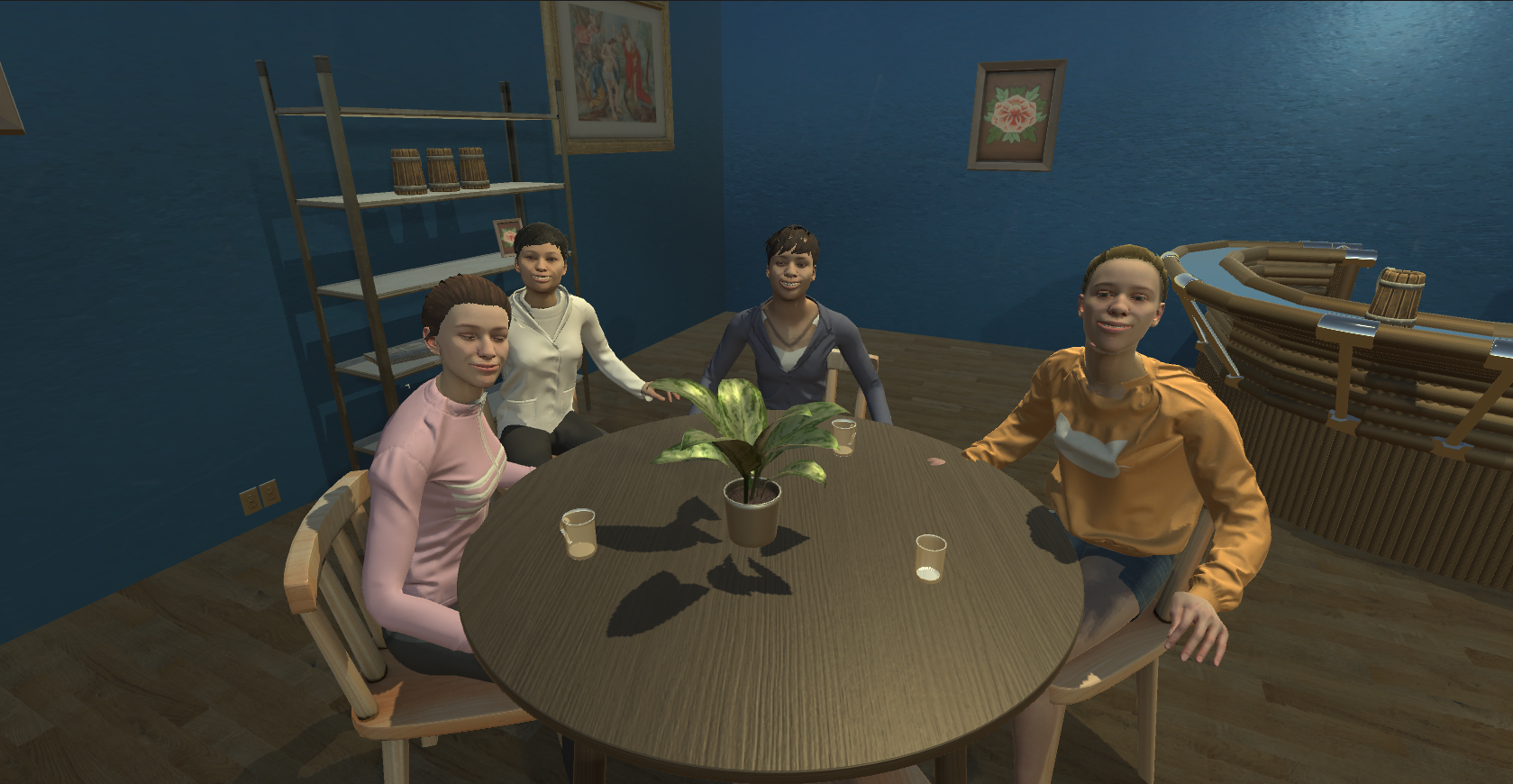}
        \label{fig:females-vhbar}
    }
    \subfigure[Agents of both genders (2 male and 2 female) are at the table.]{
        \includegraphics[width=0.47\textwidth]{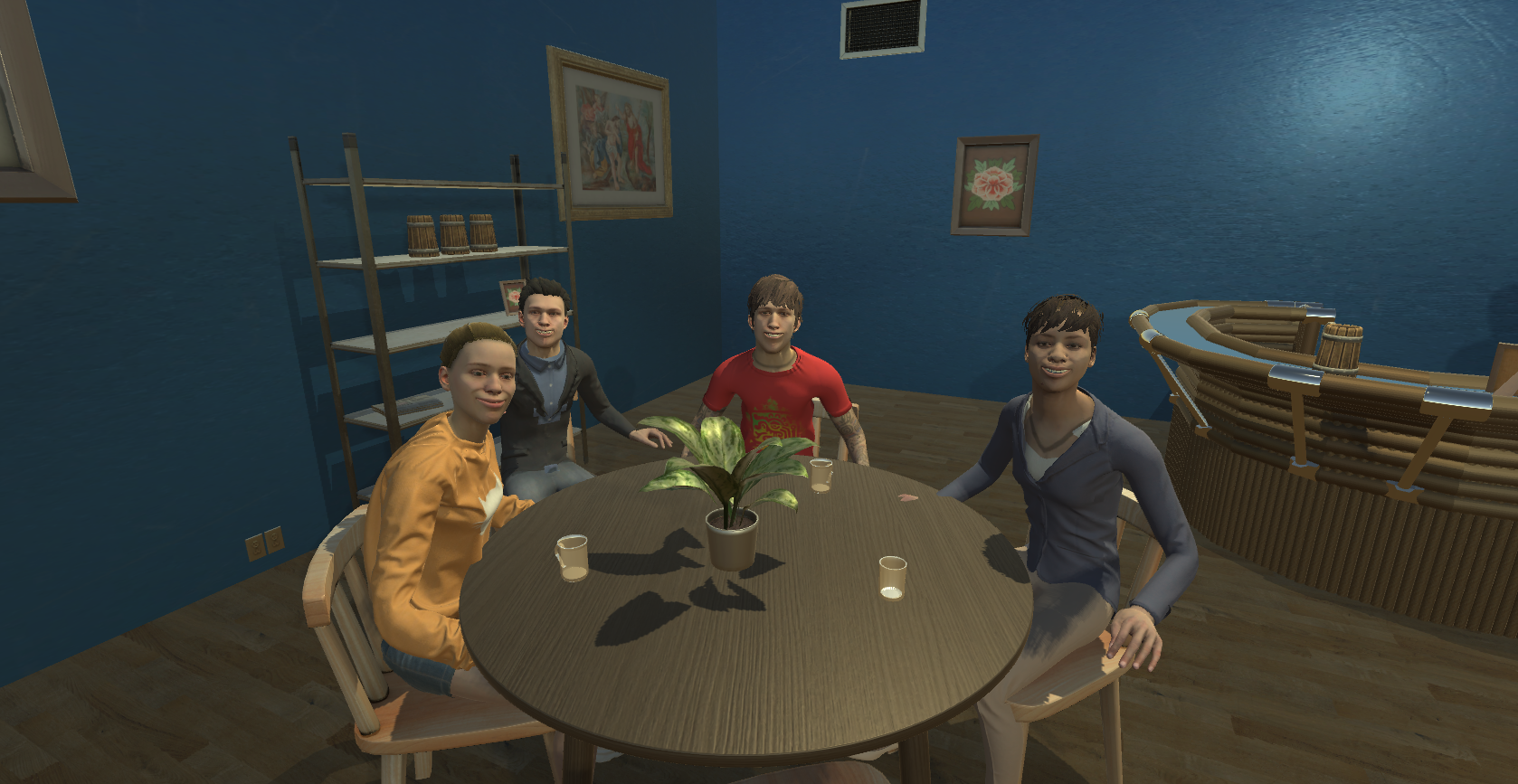}
        \label{fig:both-vhbar}
    }
    \caption{The figures present the First Person View next to the table. In these figures, all agents are expressing a positive face. Figure~\ref{fig:males-vhbar} presents a table with four male agents randomly set. Similarly, Figure~\ref{fig:females-vhbar} features a table with four female agents randomly set. Finally, Figure~\ref{fig:both-vhbar} shows agents of both genders, two male and two female, randomly set at the table.}
    \label{fig:vhbar-agents}
\end{figure*}

\subsection{Stimuli Creation}
\label{sec:stimuli}
Aiming to address our research questions, we selected eight characters (four males and four females, depicted in (a) and (b) in Figure~\ref{fig:vhbar-agents}) from Mixamo~\footnote{https://www.mixamo.com/} with a human-like appearance. These characters come with a body rig and various animations, including a sitting position. For facial animations, we constructed the facial rig using a Blender Add-on called FaceIt~\footnote{https://faceit-doc.readthedocs.io/en/latest/}. With all facial bones already configured by FaceIt, we applied an expression preset available in the Blender Add-on. The expression preset used was ARKit, the Augmented Reality development kit employed in iOS AR tools. ARKit contains a pattern with 52 blend shapes utilized for animating the face. With the facial rig ready for use, we utilized the Live Capture~\footnote{https://docs.unity3d.com/Packages/com.unity.live-capture@1.0/manual/index.html} package in the Unity Engine to record the neutral, positive, and negative expressions of four different volunteers (two men and two women). The volunteers were asked to record their expressions of happiness and anger which were represented as positive and negative emotions, respectively. This was inspired by the work of Cohn~\cite{cohn2009happiness}. The recorded expressions were then used to animate the faces of two characters of the same gender for each volunteer. The Animation Rigging package was utilized to set the view goal of the characters. For this project, we chose to use accessible and affordable tools to develop interactive simulations. Therefore, we avoided the use of state-of-the-art Metahuman technology\footnote{https://www.unrealengine.com/en-US/metahuman}, 
and we used Blender and Unity, focused on games and interactive applications.

For our studies, we created an indoor virtual bar. The participant started from a designated door, as shown in Figure~\ref{fig:vhbar-start}, where they could observe two tables—one with five chairs and the other with four chairs—and a deck table with a mixologist. The bar featured decorations such as pictures on the wall, plants, and beer glasses. Each table was set with glasses and a plant. Figures~\ref{fig:vhbar-top} and \ref{fig:inferior-right} illustrate the layout of the virtual bar.

To ensure that participants did not get distracted, only one table was equipped with VHs. Specifically, one of the tables did not have any VHs, while the other had four VHs. In the absence of interaction with the participant, the VHs always faced towards the table. However, when a user approached the table within a certain distance, the VHs turned to face the user, giving the user the feeling of being observed.

We had three types of character arrangements based on gender; the table could have only male VHs, only female VHs, or a combination of two female and two male VHs. In the latter case, both the two female and two male VH models were chosen randomly. The Figures~\ref{fig:vhbar-agents} show the participant's view next to the agents' table. As depicted in Figure~\ref{fig:vhbar-agents}, the VHs looked towards the participant when approached, aiming for participants to evaluate the expressed emotions (more details in Sections~\ref{sec:participants_action} and~\ref{sec:questionnaire}). Additionally, all VHs at the table exhibited the same emotion, which could be positive, neutral, or negative. Furthermore, VHs were randomly distributed to each chair.

To conduct this experiment, we used the following specified technologies. The Unity Engine version 2020.3.40f1~\footnote{https://unity.com/} was utilized to create and build the virtual environment, detailed in this section. The virtual environment was made accessible through WebGL, allowing participants to access it using a web browser. The WebGL application was hosted on the laboratory website. Additionally, we utilized Firebase Realtime Database~\footnote{https://firebase.google.com/} to store participants' answers during the interaction. To save their responses in the database, each participant needed to interact with all scenes presented in Section~\ref{sec:questionnaire}.


\subsection{Participant's Actions}
\label{sec:participants_action}

Based on the studies presented in the literature, which include Zojaji et al. and Zell et al.~\cite{zojaji2020influence,zell2020perception}, we provided the participant with the freedom to walk and interact with the environment. To control their movement, we used the keyboard. The participant was instructed to use the up arrow key to move forward and the down arrow key to move backward, which is a common control scheme used in games.
The participant could also utilize the mouse to rotate the avatar camera for a comprehensive view of the environment. The rotation along the \textit{X}-axis was restricted between -80 and 80 degrees. As the participant approached the VHs, a message appeared in the top left of the screen after a few seconds (following the VHs' full expressions), suggesting to the participant to press a keyboard button to exit the interaction. Upon pressing the button, the interaction concluded, and the questionnaire then appeared on the participant's screen, as detailed in Section~\ref{sec:questionnaire}


\subsection{Questionnaire}
\label{sec:questionnaire}

In Section~\ref{sec:introduction}, we presented three hypotheses. To test these hypotheses, we conducted an online survey using the Qualtrics survey tool and a virtual environment created with Unity and WebGL. The survey comprised four parts: a consent form, demography questions, the interactive scenario,
and final comments. The consent form explained the research's purpose and requested participants' agreement. Demography questions sought personal information, including gender, education, age (average), and familiarity with technology and Computer Graphics. Please note that we have omitted information about the project due to blind submission. These two steps mentioned were presented through a Qualtrics link. After that, still using Qualtrics, participants received a link to the interactive environment.

In the interactive scenario part, participants received a tutorial on how to interact with VHs in the experiment and 
how to return to Qualtrics to the final comments part.
The context of the interaction was presented to the participant through text before presenting the interactive environment. The text was as follows:``The participant was in a bar with friends. However, they realized the wallet was forgotten after leaving the bar. Upon re-entering the bar, the participant discovered that the table was now occupied by other people (the VHs). Therefore, the participant's goal was to approach the table and look for their 
wallet.´´ We chose this specific bar context based on insights from related studies regarding group perceptions (Section~\ref{sec:related_work}), as it represents an environment with social group relations.

All participants interacted in the virtual environment nine different times,
in a combination of three types of emotions and three gender distributions of VHs, as presented in Section~\ref{sec:stimuli}. 
For every interaction, participants answered five questions, as specified in Table~\ref{tab:questions}. Questions $Q1$ and $Q4$ addressed $H0_2$, examining whether individuals felt similarly comfortable with groups of VHs featuring different genders and emotions. Questions $Q2$ and $Q3$ pertained to $H0_1$, investigating whether individuals felt similarly comfortable with groups of VHs featuring different genders and emotions. Question $Q1$ was also associated with $H0_3$, aiming to assess whether individuals perceived realism similarly across different groups of VHs. Finally, at the end of the survey, an optional text box was provided for participants to share any relevant thoughts or comments.

\begin{table*}[!htb]
    \centering
    \begin{tabular}{|c|c|c|}
    \hline
    \textbf{Hypothesis} & \textbf{Related Question(s)}                                                                                                                                                                                         & \textbf{Possible Answers}                                                                                                                       \\ \hline
    $H0_1$              & \begin{tabular}[c]{@{}c@{}}Q2: How did you classify the emotions \\ of these characters in this scene? \\ \\ Q3: How did you classify the intensity \\ of the emotions of the characters in this scene?\end{tabular} & \begin{tabular}[c]{@{}c@{}}Negative; Neutral;\\ Positive; I don't know\\ \\ Weak; Neutral;\\ Strong; I don't know\end{tabular}                  \\ \hline
    $H0_2$              & \begin{tabular}[c]{@{}c@{}}Q1: How did you feel in this scene?\\ \\ \\ Q4: How much did you feel comfort in this scene?\end{tabular}                                                                                 & \begin{tabular}[c]{@{}c@{}}Negative; Neutral;\\ Positive; I don't know\\ \\ 5-Likert from very uncomfortable\\ to very comfortable\end{tabular} \\ \hline
    $H0_3$              & Q5: How much was this scene realistic?                                                                                                                                                                               & \begin{tabular}[c]{@{}c@{}}5-Likert from very unrealistic\\ to very realistic\end{tabular}                                                      \\ \hline
    \end{tabular}
    \caption{The table presents the questions applied in the questionnaire to the participants regarding their perception of VHs. The 'Hypothesis' column outlines the hypothesis the question aims to address, while the 'Question' column provides the actual questions used in the questionnaire. The 'Possible Answers' column lists the options available for participants to choose from.}
    \label{tab:questions}
\end{table*}

\section{Results}
\label{sec:results}
In this section, we present the main findings of the research conducted in this work. A questionnaire was distributed on social media platforms, and 60 individuals volunteered to participate. Out of these volunteers, 40 (66.6\%) were male, while 20 (33.4\%) were female. A majority of the participants (68.3\%) were aged between 18 and 20 years. Additionally, 30 (50.0\%) participants had completed their secondary education. Out of all the participants, 31 individuals claimed to have high experience with technology, while 36 individuals reported having little or no experience with computer graphics. 

The statistical analyses performed were a factorial ANOVA 2 (Participant Gender - Men and Women) x 3 (VH Gender - Male, Female, Both) x 3 (VH Emotion - Negative, Neutral, Positive), considering a significance level of 5\%. We conducted two types of post-hoc test to examine further the significant effects observed in the analysis. First of all, we conducted a Bartlett test to measure the equal variances (or not). In case of equal variances, we used the Tukey HSD test. In cases where the groups of the data did not have equal variances, we used the \textit{Tamhane} test.

\subsection{Analysis of Emotion Recognition}
\label{sec:vh_emotion}


To evaluate the emotions of the participants in the questionnaire, we assigned numerical values to the answer options for $Q2$ and $Q3$. For $Q2$, a score of 1 was assigned for a negative response, 2 for a neutral response, 3 for a positive response, and 0 for 'I don't know' responses. Similarly, for $Q3$, we assigned a value of 1 for a weak emotion response, 2 for a neutral emotion response, 3 for a strong emotion response, and 0 for 'I don't know' responses. In other words, the participant would recognize correctly the VH's positive emotion if they chose the ``Positive" option as the answer to $Q2$, and similarly for the negative emotion and neutral. These values were utilized for statistical analysis.

In relation to emotion recognition, starting with $H0_1$, we identified two significant results in the statistical test using the answers of $Q2$. Firstly, there was a main effect of VH Emotion (\textit{F}(2, 537) = $138.93$, \textit{p}$<.001$), and secondly, an interaction effect between VH Gender and VH Emotion (\textit{F}(4, 535) = $4.87$, \textit{p}$<.001$). Upon noting that the data related to VH Emotion did not exhibit equal variance (Stats = $13.96$, \textit{p}$<.001$), we conducted post-hoc tests, revealing significant differences when comparing negative emotion between the other conditions (neutral - \textit{p}$<.001$; positive - \textit{p}$<.001$), and also a significant difference between neutral and positive emotions (\textit{p}$<.001$). Observing the average scores for emotion recognition in Table~\ref{tab:vh-emotion}, associated with $Q2$, we observe that, in most cases, participants correctly identified negative, neutral, and positive emotions.
Therefore, \textbf{due to to the fact that the recognition score was close to 1 (negative response) for negative emotion (average of $1.55$), close to 2 (neutral response) for neutral emotion (average of $1.96$), and close to 3 (positive response) for positive emotion (average of $2.74$), we can say that the participants correctly identified the emotions expressed by the groups of VHs.} 



Still in relation to $H0_1$, focusing on the interaction between the VH Gender and the VH Emotion, first of all, the data did not have equal variance (Stats = $48.95$ and $p<.001$). The results of the post-hoc test presented significant results, comparing the negative male emotion with the negative female emotion ($p<.001$). We also found a significant result comparing the positive male emotion and the others male emotions (neutral - $p<.001$; negative - $p<.001$), comparing the positive female and the others female emotions (neutral - $p<.001$; negative - $p<.001$), and comparing the positive emotion of both gender condition (a table with female and male VHs) and the other emotions (neutral - $p<.001$; negative - $p<.001$). 
Based on our results, it appears \textbf{that individuals are better at recognizing negative emotions in groups that only had female VHs than groups that only had male VHs. The average recognition score for negative female emotions was $1.20$, which was closer to a perfect score of 1, while the average recognition score for negative male emotions was 1.86, indicating a less accurate perception. On the other hand, We obtained similar results when we varied the VH gender condition for negative and neutral emotions. In this case, people's ability to recognize emotions did not differ significantly.
}

\begin{table*}[!htb]
    \centering
    \begin{tabular}{|ccccccccc|}
    \hline
    \multicolumn{9}{|c|}{\textbf{VH Emotion Answers}}                                                                                                                                                                                                                                                                                                                                                                                \\ \hline
    \multicolumn{1}{|c|}{\multirow{2}{*}{\textbf{Participant Gender}}} & \multicolumn{2}{c|}{\textbf{User's Answer}}                                             & \multicolumn{1}{c|}{\multirow{2}{*}{\textbf{VH Emotion}}} & \multicolumn{2}{c|}{\textbf{User's Answer}}                                             & \multicolumn{1}{c|}{\multirow{2}{*}{\textbf{VH Gender}}} & \multicolumn{2}{c|}{\textbf{User's Answer}}      \\ \cline{2-3} \cline{5-6} \cline{8-9} 
    \multicolumn{1}{|c|}{}                                             & \multicolumn{1}{c|}{\textbf{Avg}}          & \multicolumn{1}{c|}{\textbf{STD}}          & \multicolumn{1}{c|}{}                                     & \multicolumn{1}{c|}{\textbf{Avg}}          & \multicolumn{1}{c|}{\textbf{STD}}          & \multicolumn{1}{c|}{}                                    & \multicolumn{1}{c|}{\textbf{Avg}} & \textbf{STD} \\ \hline
    \multicolumn{1}{|c|}{\multirow{9}{*}{Male}}                        & \multicolumn{1}{c|}{\multirow{9}{*}{2.07}} & \multicolumn{1}{c|}{\multirow{9}{*}{0.83}} & \multicolumn{1}{c|}{\multirow{3}{*}{Negative}}            & \multicolumn{1}{c|}{\multirow{3}{*}{1.54}} & \multicolumn{1}{c|}{\multirow{3}{*}{0.78}} & \multicolumn{1}{c|}{Male}                                & \multicolumn{1}{c|}{1.90}         & 0.90         \\ \cline{7-9} 
    \multicolumn{1}{|c|}{}                                             & \multicolumn{1}{c|}{}                      & \multicolumn{1}{c|}{}                      & \multicolumn{1}{c|}{}                                     & \multicolumn{1}{c|}{}                      & \multicolumn{1}{c|}{}                      & \multicolumn{1}{c|}{Female}                              & \multicolumn{1}{c|}{1.15}         & 0.42         \\ \cline{7-9} 
    \multicolumn{1}{|c|}{}                                             & \multicolumn{1}{c|}{}                      & \multicolumn{1}{c|}{}                      & \multicolumn{1}{c|}{}                                     & \multicolumn{1}{c|}{}                      & \multicolumn{1}{c|}{}                      & \multicolumn{1}{c|}{Both}                                & \multicolumn{1}{c|}{1.57}         & 0.78         \\ \cline{4-9} 
    \multicolumn{1}{|c|}{}                                             & \multicolumn{1}{c|}{}                      & \multicolumn{1}{c|}{}                      & \multicolumn{1}{c|}{\multirow{3}{*}{Neutral}}             & \multicolumn{1}{c|}{\multirow{3}{*}{1.95}} & \multicolumn{1}{c|}{\multirow{3}{*}{0.59}} & \multicolumn{1}{c|}{Male}                                & \multicolumn{1}{c|}{1.90}         & 0.70         \\ \cline{7-9} 
    \multicolumn{1}{|c|}{}                                             & \multicolumn{1}{c|}{}                      & \multicolumn{1}{c|}{}                      & \multicolumn{1}{c|}{}                                     & \multicolumn{1}{c|}{}                      & \multicolumn{1}{c|}{}                      & \multicolumn{1}{c|}{Female}                              & \multicolumn{1}{c|}{2.07}         & 0.41         \\ \cline{7-9} 
    \multicolumn{1}{|c|}{}                                             & \multicolumn{1}{c|}{}                      & \multicolumn{1}{c|}{}                      & \multicolumn{1}{c|}{}                                     & \multicolumn{1}{c|}{}                      & \multicolumn{1}{c|}{}                      & \multicolumn{1}{c|}{Both}                                & \multicolumn{1}{c|}{1.87}         & 0.60         \\ \cline{4-9} 
    \multicolumn{1}{|c|}{}                                             & \multicolumn{1}{c|}{}                      & \multicolumn{1}{c|}{}                      & \multicolumn{1}{c|}{\multirow{3}{*}{Positive}}            & \multicolumn{1}{c|}{\multirow{3}{*}{2.73}} & \multicolumn{1}{c|}{\multirow{3}{*}{0.63}} & \multicolumn{1}{c|}{Male}                                & \multicolumn{1}{c|}{2.80}         & 0.51         \\ \cline{7-9} 
    \multicolumn{1}{|c|}{}                                             & \multicolumn{1}{c|}{}                      & \multicolumn{1}{c|}{}                      & \multicolumn{1}{c|}{}                                     & \multicolumn{1}{c|}{}                      & \multicolumn{1}{c|}{}                      & \multicolumn{1}{c|}{Female}                              & \multicolumn{1}{c|}{2.72}         & 0.71         \\ \cline{7-9} 
    \multicolumn{1}{|c|}{}                                             & \multicolumn{1}{c|}{}                      & \multicolumn{1}{c|}{}                      & \multicolumn{1}{c|}{}                                     & \multicolumn{1}{c|}{}                      & \multicolumn{1}{c|}{}                      & \multicolumn{1}{c|}{Both}                                & \multicolumn{1}{c|}{2.67}         & 0.65         \\ \hline
    \multicolumn{1}{|c|}{\multirow{9}{*}{Female}}                      & \multicolumn{1}{c|}{\multirow{9}{*}{2.07}} & \multicolumn{1}{c|}{\multirow{9}{*}{0.88}} & \multicolumn{1}{c|}{\multirow{3}{*}{Negative}}            & \multicolumn{1}{c|}{\multirow{3}{*}{1.48}} & \multicolumn{1}{c|}{\multirow{3}{*}{0.79}} & \multicolumn{1}{c|}{Male}                                & \multicolumn{1}{c|}{1.80}         & 1.05         \\ \cline{7-9} 
    \multicolumn{1}{|c|}{}                                             & \multicolumn{1}{c|}{}                      & \multicolumn{1}{c|}{}                      & \multicolumn{1}{c|}{}                                     & \multicolumn{1}{c|}{}                      & \multicolumn{1}{c|}{}                      & \multicolumn{1}{c|}{Female}                              & \multicolumn{1}{c|}{1.30}         & 0.57         \\ \cline{7-9} 
    \multicolumn{1}{|c|}{}                                             & \multicolumn{1}{c|}{}                      & \multicolumn{1}{c|}{}                      & \multicolumn{1}{c|}{}                                     & \multicolumn{1}{c|}{}                      & \multicolumn{1}{c|}{}                      & \multicolumn{1}{c|}{Both}                                & \multicolumn{1}{c|}{1.35}         & 0.58         \\ \cline{4-9} 
    \multicolumn{1}{|c|}{}                                             & \multicolumn{1}{c|}{}                      & \multicolumn{1}{c|}{}                      & \multicolumn{1}{c|}{\multirow{3}{*}{Neutral}}             & \multicolumn{1}{c|}{\multirow{3}{*}{1.98}} & \multicolumn{1}{c|}{\multirow{3}{*}{0.62}} & \multicolumn{1}{c|}{Male}                                & \multicolumn{1}{c|}{2.00}         & 0.72         \\ \cline{7-9} 
    \multicolumn{1}{|c|}{}                                             & \multicolumn{1}{c|}{}                      & \multicolumn{1}{c|}{}                      & \multicolumn{1}{c|}{}                                     & \multicolumn{1}{c|}{}                      & \multicolumn{1}{c|}{}                      & \multicolumn{1}{c|}{Female}                              & \multicolumn{1}{c|}{2.00}         & 0.45         \\ \cline{7-9} 
    \multicolumn{1}{|c|}{}                                             & \multicolumn{1}{c|}{}                      & \multicolumn{1}{c|}{}                      & \multicolumn{1}{c|}{}                                     & \multicolumn{1}{c|}{}                      & \multicolumn{1}{c|}{}                      & \multicolumn{1}{c|}{Both}                                & \multicolumn{1}{c|}{1.95}         & 0.68         \\ \cline{4-9} 
    \multicolumn{1}{|c|}{}                                             & \multicolumn{1}{c|}{}                      & \multicolumn{1}{c|}{}                      & \multicolumn{1}{c|}{\multirow{3}{*}{Positive}}            & \multicolumn{1}{c|}{\multirow{3}{*}{2.76}} & \multicolumn{1}{c|}{\multirow{3}{*}{0.69}} & \multicolumn{1}{c|}{Male}                                & \multicolumn{1}{c|}{2.65}         & 0.87         \\ \cline{7-9} 
    \multicolumn{1}{|c|}{}                                             & \multicolumn{1}{c|}{}                      & \multicolumn{1}{c|}{}                      & \multicolumn{1}{c|}{}                                     & \multicolumn{1}{c|}{}                      & \multicolumn{1}{c|}{}                      & \multicolumn{1}{c|}{Female}                              & \multicolumn{1}{c|}{2.85}         & 0.67         \\ \cline{7-9} 
    \multicolumn{1}{|c|}{}                                             & \multicolumn{1}{c|}{}                      & \multicolumn{1}{c|}{}                      & \multicolumn{1}{c|}{}                                     & \multicolumn{1}{c|}{}                      & \multicolumn{1}{c|}{}                      & \multicolumn{1}{c|}{Both}                                & \multicolumn{1}{c|}{2.80}         & 0.52         \\ \hline
    \end{tabular}
    \caption{Average and standard deviation of the VH Emotion answers (Q2 in the questionnaire following Table~\ref{tab:questions}). For $Q2$ answer options, we assigned values to have an emotion recognition score value, as follows: 1 for negative, 2 for neutral, 3 for positive, and 0 for 'I don't know' responses.}
    \label{tab:vh-emotion}
\end{table*}

Continuing to analyze $H0_1$, we found only one significant result, using $Q3$ (emotion intensity score), 
in the main effect of VH Emotion (F(2, 537) = $53.62$, $p<.001$; With no equal variance - Stats = 41.16, $p<.001$). Following the post-hoc analysis, we identified significant differences when comparing negative emotion with the other two conditions (neutral - $p=.001$; positive - $p<.001$) and between neutral and positive emotions ($p<.001$). As we can see in Table~\ref{tab:vh-emotion-intensity},
most responses for VH neutral emotion were classified as neutral intensity, positive emotion responses were predominantly strong intensity, and negative emotion responses were a mix between neutral and strong intensities. Therefore, \textbf{we can conclude that people perceived more intensity in positive emotion than in negative emotion (with general averages of $2.75$ and $2.27$, respectively), positive emotion had more intensity than neutral emotion (with general averages of $2.75$ and $2.00$, respectively), and negative emotion had more intensity than neutral emotion (with general averages of $2.27$ and $2.00$, respectively).}
It can be concluded that \textbf{we can reject $H0_1$, as the results showed that people accurately recognized the emotions expressed by groups of VHs, and people were more accurate in recognizing negative emotion when the group of VHs was all-female than when the group of VHs was all-male.}


\begin{table*}[!htb]
    \centering
    \begin{tabular}{|ccccccccc|}
    \hline
    \multicolumn{9}{|c|}{\textbf{Intensity of VH Emotion Answers}}                                                                                                                                                                                                                                                                                                                                                                   \\ \hline
    \multicolumn{1}{|c|}{\multirow{2}{*}{\textbf{Participant Gender}}} & \multicolumn{2}{c|}{\textbf{User's Answer}}                                             & \multicolumn{1}{c|}{\multirow{2}{*}{\textbf{VH Emotion}}} & \multicolumn{2}{c|}{\textbf{User's Answer}}                                             & \multicolumn{1}{c|}{\multirow{2}{*}{\textbf{VH Gender}}} & \multicolumn{2}{c|}{\textbf{User's Answer}}      \\ \cline{2-3} \cline{5-6} \cline{8-9} 
    \multicolumn{1}{|c|}{}                                             & \multicolumn{1}{c|}{\textbf{Avg}}          & \multicolumn{1}{c|}{\textbf{STD}}          & \multicolumn{1}{c|}{}                                     & \multicolumn{1}{c|}{\textbf{Avg}}          & \multicolumn{1}{c|}{\textbf{STD}}          & \multicolumn{1}{c|}{}                                    & \multicolumn{1}{c|}{\textbf{Avg}} & \textbf{STD} \\ \hline
    \multicolumn{1}{|c|}{\multirow{9}{*}{Male}}                        & \multicolumn{1}{c|}{\multirow{9}{*}{2.35}} & \multicolumn{1}{c|}{\multirow{9}{*}{0.73}} & \multicolumn{1}{c|}{\multirow{3}{*}{Negative}}            & \multicolumn{1}{c|}{\multirow{3}{*}{2.32}} & \multicolumn{1}{c|}{\multirow{3}{*}{0.77}} & \multicolumn{1}{c|}{Male}                                & \multicolumn{1}{c|}{2.32}         & 0.76         \\ \cline{7-9} 
    \multicolumn{1}{|c|}{}                                             & \multicolumn{1}{c|}{}                      & \multicolumn{1}{c|}{}                      & \multicolumn{1}{c|}{}                                     & \multicolumn{1}{c|}{}                      & \multicolumn{1}{c|}{}                      & \multicolumn{1}{c|}{Female}                              & \multicolumn{1}{c|}{2.32}         & 0.88         \\ \cline{7-9} 
    \multicolumn{1}{|c|}{}                                             & \multicolumn{1}{c|}{}                      & \multicolumn{1}{c|}{}                      & \multicolumn{1}{c|}{}                                     & \multicolumn{1}{c|}{}                      & \multicolumn{1}{c|}{}                      & \multicolumn{1}{c|}{Both}                                & \multicolumn{1}{c|}{2.32}         & 0.69         \\ \cline{4-9} 
    \multicolumn{1}{|c|}{}                                             & \multicolumn{1}{c|}{}                      & \multicolumn{1}{c|}{}                      & \multicolumn{1}{c|}{\multirow{3}{*}{Neutral}}             & \multicolumn{1}{c|}{\multirow{3}{*}{2.00}} & \multicolumn{1}{c|}{\multirow{3}{*}{0.69}} & \multicolumn{1}{c|}{Male}                                & \multicolumn{1}{c|}{1.87}         & 0.75         \\ \cline{7-9} 
    \multicolumn{1}{|c|}{}                                             & \multicolumn{1}{c|}{}                      & \multicolumn{1}{c|}{}                      & \multicolumn{1}{c|}{}                                     & \multicolumn{1}{c|}{}                      & \multicolumn{1}{c|}{}                      & \multicolumn{1}{c|}{Female}                              & \multicolumn{1}{c|}{2.05}         & 0.59         \\ \cline{7-9} 
    \multicolumn{1}{|c|}{}                                             & \multicolumn{1}{c|}{}                      & \multicolumn{1}{c|}{}                      & \multicolumn{1}{c|}{}                                     & \multicolumn{1}{c|}{}                      & \multicolumn{1}{c|}{}                      & \multicolumn{1}{c|}{Both}                                & \multicolumn{1}{c|}{2.10}         & 0.70         \\ \cline{4-9} 
    \multicolumn{1}{|c|}{}                                             & \multicolumn{1}{c|}{}                      & \multicolumn{1}{c|}{}                      & \multicolumn{1}{c|}{\multirow{3}{*}{Positive}}            & \multicolumn{1}{c|}{\multirow{3}{*}{2.73}} & \multicolumn{1}{c|}{\multirow{3}{*}{0.54}} & \multicolumn{1}{c|}{Male}                                & \multicolumn{1}{c|}{2.75}         & 0.49         \\ \cline{7-9} 
    \multicolumn{1}{|c|}{}                                             & \multicolumn{1}{c|}{}                      & \multicolumn{1}{c|}{}                      & \multicolumn{1}{c|}{}                                     & \multicolumn{1}{c|}{}                      & \multicolumn{1}{c|}{}                      & \multicolumn{1}{c|}{Female}                              & \multicolumn{1}{c|}{2.67}         & 0.61         \\ \cline{7-9} 
    \multicolumn{1}{|c|}{}                                             & \multicolumn{1}{c|}{}                      & \multicolumn{1}{c|}{}                      & \multicolumn{1}{c|}{}                                     & \multicolumn{1}{c|}{}                      & \multicolumn{1}{c|}{}                      & \multicolumn{1}{c|}{Both}                                & \multicolumn{1}{c|}{2.77}         & 0.53         \\ \hline
    \multicolumn{1}{|c|}{\multirow{9}{*}{Female}}                      & \multicolumn{1}{c|}{\multirow{9}{*}{2.32}} & \multicolumn{1}{c|}{\multirow{9}{*}{0.74}} & \multicolumn{1}{c|}{\multirow{3}{*}{Negative}}            & \multicolumn{1}{c|}{\multirow{3}{*}{2.18}} & \multicolumn{1}{c|}{\multirow{3}{*}{0.89}} & \multicolumn{1}{c|}{Male}                                & \multicolumn{1}{c|}{2.35}         & 0.87         \\ \cline{7-9} 
    \multicolumn{1}{|c|}{}                                             & \multicolumn{1}{c|}{}                      & \multicolumn{1}{c|}{}                      & \multicolumn{1}{c|}{}                                     & \multicolumn{1}{c|}{}                      & \multicolumn{1}{c|}{}                      & \multicolumn{1}{c|}{Female}                              & \multicolumn{1}{c|}{2.15}         & 0.81         \\ \cline{7-9} 
    \multicolumn{1}{|c|}{}                                             & \multicolumn{1}{c|}{}                      & \multicolumn{1}{c|}{}                      & \multicolumn{1}{c|}{}                                     & \multicolumn{1}{c|}{}                      & \multicolumn{1}{c|}{}                      & \multicolumn{1}{c|}{Both}                                & \multicolumn{1}{c|}{2.05}         & 0.99         \\ \cline{4-9} 
    \multicolumn{1}{|c|}{}                                             & \multicolumn{1}{c|}{}                      & \multicolumn{1}{c|}{}                      & \multicolumn{1}{c|}{\multirow{3}{*}{Neutral}}             & \multicolumn{1}{c|}{\multirow{3}{*}{1.98}} & \multicolumn{1}{c|}{\multirow{3}{*}{0.59}} & \multicolumn{1}{c|}{Male}                                & \multicolumn{1}{c|}{1.95}         & 0.60         \\ \cline{7-9} 
    \multicolumn{1}{|c|}{}                                             & \multicolumn{1}{c|}{}                      & \multicolumn{1}{c|}{}                      & \multicolumn{1}{c|}{}                                     & \multicolumn{1}{c|}{}                      & \multicolumn{1}{c|}{}                      & \multicolumn{1}{c|}{Female}                              & \multicolumn{1}{c|}{2.00}         & 0.64         \\ \cline{7-9} 
    \multicolumn{1}{|c|}{}                                             & \multicolumn{1}{c|}{}                      & \multicolumn{1}{c|}{}                      & \multicolumn{1}{c|}{}                                     & \multicolumn{1}{c|}{}                      & \multicolumn{1}{c|}{}                      & \multicolumn{1}{c|}{Both}                                & \multicolumn{1}{c|}{2.00}         & 0.56         \\ \cline{4-9} 
    \multicolumn{1}{|c|}{}                                             & \multicolumn{1}{c|}{}                      & \multicolumn{1}{c|}{}                      & \multicolumn{1}{c|}{\multirow{3}{*}{Positive}}            & \multicolumn{1}{c|}{\multirow{3}{*}{2.80}} & \multicolumn{1}{c|}{\multirow{3}{*}{0.40}} & \multicolumn{1}{c|}{Male}                                & \multicolumn{1}{c|}{2.90}         & 0.30         \\ \cline{7-9} 
    \multicolumn{1}{|c|}{}                                             & \multicolumn{1}{c|}{}                      & \multicolumn{1}{c|}{}                      & \multicolumn{1}{c|}{}                                     & \multicolumn{1}{c|}{}                      & \multicolumn{1}{c|}{}                      & \multicolumn{1}{c|}{Female}                              & \multicolumn{1}{c|}{2.75}         & 0.44         \\ \cline{7-9} 
    \multicolumn{1}{|c|}{}                                             & \multicolumn{1}{c|}{}                      & \multicolumn{1}{c|}{}                      & \multicolumn{1}{c|}{}                                     & \multicolumn{1}{c|}{}                      & \multicolumn{1}{c|}{}                      & \multicolumn{1}{c|}{Both}                                & \multicolumn{1}{c|}{2.75}         & 0.44         \\ \hline
    \end{tabular}
    \caption{Average and standard deviation of the Intensity of VH Emotion answers (Q3 in the questionnaire following Table~\ref{tab:questions}). For $Q3$ answer options, we assigned values of 1 for weak, 2 for neutral, 3 for strong, and 0 for 'I don't know' responses.}
    \label{tab:vh-emotion-intensity}
\end{table*}

\subsection{Analysis of Comfort Perception}
\label{sec:comfort}


For statistics analysis of $H0_2$, we focused on $Q1$ and $Q4$ as outlined in Table~\ref{tab:questions}. For $Q1$, to have a score on what the participant felt emotionally during the interaction, we attributed the following values to the next answers: 1 to negative, 2 to neutral, 3 to positive, and finally, 0 to I don't know. For $Q4$, to have a comfort score (that is, the level of comfort felt during the interaction), we followed the 5-Likert scale where very uncomfortable is 1 and very comfortable is 5.

In the statistical analysis of the $Q1$ question (Table~\ref{tab:questions}), we identified two significant results. The first was the main effect of the VH Emotion (F(2, 537) = $17.54$ and $p<.001$; No equal variance (Stats = $16.89$ and $p<.001$)). The second was an interaction effect between the VH Gender and the Participant Gender (F(2,547) = $3.11$ and $p=.045$; Bartlett Result - Stats = $10.36$ and $p=.065$), but we did not find significant results in the paired comparisons resulted by the post-hoc. Because of that, only observing the main effect of VH Emotion, after the post-hoc, we found significant results comparing the negative emotion and the other conditions (neutral - $p<.001$; positive - $p<.001$). As we can see in Table~\ref{tab:user-feeling}, most of the positive feelings responded by participants were associated with positive and neutral emotions expressed by groups of VHs, while negative emotions were more associated with neutral feelings. 
Therefore, \textbf{we can say that the people felt better with positive emotion than negative emotion (respectively, general averages of $2.21$ and $1.79$), felt better with neutral than with negative emotion (respectively, general averages of $2.16$ and $1.79$), and similar between neutral and positive.} 

\begin{table*}[!htb]
    \centering
    \begin{tabular}{|ccccccccc|}
    \hline
    \multicolumn{9}{|c|}{\textbf{User`s Feeling Answers}}                                                                                                                                                                                                                                                                                                                                                                   \\ \hline
    \multicolumn{1}{|c|}{\multirow{2}{*}{\textbf{Participant Gender}}} & \multicolumn{2}{c|}{\textbf{User's Answer}}                                             & \multicolumn{1}{c|}{\multirow{2}{*}{\textbf{VH Emotion}}} & \multicolumn{2}{c|}{\textbf{User's Answer}}                                             & \multicolumn{1}{c|}{\multirow{2}{*}{\textbf{VH Gender}}} & \multicolumn{2}{c|}{\textbf{User's Answer}}      \\ \cline{2-3} \cline{5-6} \cline{8-9} 
    \multicolumn{1}{|c|}{}                                    & \multicolumn{1}{c|}{\textbf{Avg}}          & \multicolumn{1}{c|}{\textbf{STD}}          & \multicolumn{1}{c|}{}                                     & \multicolumn{1}{c|}{\textbf{Avg}}          & \multicolumn{1}{c|}{\textbf{STD}}          & \multicolumn{1}{c|}{}                                    & \multicolumn{1}{c|}{\textbf{Avg}} & \textbf{STD} \\ \hline
    \multicolumn{1}{|c|}{\multirow{9}{*}{Male}}               & \multicolumn{1}{c|}{\multirow{9}{*}{2.06}} & \multicolumn{1}{c|}{\multirow{9}{*}{0.75}} & \multicolumn{1}{c|}{\multirow{3}{*}{Negative}}            & \multicolumn{1}{c|}{\multirow{3}{*}{1.85}} & \multicolumn{1}{c|}{\multirow{3}{*}{0.70}} & \multicolumn{1}{c|}{Male}                                & \multicolumn{1}{c|}{1.92}         & 0.79         \\ \cline{7-9} 
    \multicolumn{1}{|c|}{}                                    & \multicolumn{1}{c|}{}                      & \multicolumn{1}{c|}{}                      & \multicolumn{1}{c|}{}                                     & \multicolumn{1}{c|}{}                      & \multicolumn{1}{c|}{}                      & \multicolumn{1}{c|}{Female}                              & \multicolumn{1}{c|}{1.65}         & 0.62         \\ \cline{7-9} 
    \multicolumn{1}{|c|}{}                                    & \multicolumn{1}{c|}{}                      & \multicolumn{1}{c|}{}                      & \multicolumn{1}{c|}{}                                     & \multicolumn{1}{c|}{}                      & \multicolumn{1}{c|}{}                      & \multicolumn{1}{c|}{Both}                                & \multicolumn{1}{c|}{1.97}         & 0.65         \\ \cline{4-9} 
    \multicolumn{1}{|c|}{}                                    & \multicolumn{1}{c|}{}                      & \multicolumn{1}{c|}{}                      & \multicolumn{1}{c|}{\multirow{3}{*}{Neutral}}             & \multicolumn{1}{c|}{\multirow{3}{*}{2.18}} & \multicolumn{1}{c|}{\multirow{3}{*}{0.60}} & \multicolumn{1}{c|}{Male}                                & \multicolumn{1}{c|}{2.20}         & 0.64         \\ \cline{7-9} 
    \multicolumn{1}{|c|}{}                                    & \multicolumn{1}{c|}{}                      & \multicolumn{1}{c|}{}                      & \multicolumn{1}{c|}{}                                     & \multicolumn{1}{c|}{}                      & \multicolumn{1}{c|}{}                      & \multicolumn{1}{c|}{Female}                              & \multicolumn{1}{c|}{2.15}         & 0.62         \\ \cline{7-9} 
    \multicolumn{1}{|c|}{}                                    & \multicolumn{1}{c|}{}                      & \multicolumn{1}{c|}{}                      & \multicolumn{1}{c|}{}                                     & \multicolumn{1}{c|}{}                      & \multicolumn{1}{c|}{}                      & \multicolumn{1}{c|}{Both}                                & \multicolumn{1}{c|}{2.20}         & 0.56         \\ \cline{4-9} 
    \multicolumn{1}{|c|}{}                                    & \multicolumn{1}{c|}{}                      & \multicolumn{1}{c|}{}                      & \multicolumn{1}{c|}{\multirow{3}{*}{Positive}}            & \multicolumn{1}{c|}{\multirow{3}{*}{2.16}} & \multicolumn{1}{c|}{\multirow{3}{*}{0.87}} & \multicolumn{1}{c|}{Male}                                & \multicolumn{1}{c|}{2.07}         & 0.97         \\ \cline{7-9} 
    \multicolumn{1}{|c|}{}                                    & \multicolumn{1}{c|}{}                      & \multicolumn{1}{c|}{}                      & \multicolumn{1}{c|}{}                                     & \multicolumn{1}{c|}{}                      & \multicolumn{1}{c|}{}                      & \multicolumn{1}{c|}{Female}                              & \multicolumn{1}{c|}{2.17}         & 0.87         \\ \cline{7-9} 
    \multicolumn{1}{|c|}{}                                    & \multicolumn{1}{c|}{}                      & \multicolumn{1}{c|}{}                      & \multicolumn{1}{c|}{}                                     & \multicolumn{1}{c|}{}                      & \multicolumn{1}{c|}{}                      & \multicolumn{1}{c|}{Both}                                & \multicolumn{1}{c|}{2.25}         & 0.77         \\ \hline
    \multicolumn{1}{|c|}{\multirow{9}{*}{Female}}             & \multicolumn{1}{c|}{\multirow{9}{*}{2.03}} & \multicolumn{1}{c|}{\multirow{9}{*}{0.81}} & \multicolumn{1}{c|}{\multirow{3}{*}{Negative}}            & \multicolumn{1}{c|}{\multirow{3}{*}{1.68}} & \multicolumn{1}{c|}{\multirow{3}{*}{0.79}} & \multicolumn{1}{c|}{Male}                                & \multicolumn{1}{c|}{1.65}         & 0.93         \\ \cline{7-9} 
    \multicolumn{1}{|c|}{}                                    & \multicolumn{1}{c|}{}                      & \multicolumn{1}{c|}{}                      & \multicolumn{1}{c|}{}                                     & \multicolumn{1}{c|}{}                      & \multicolumn{1}{c|}{}                      & \multicolumn{1}{c|}{Female}                              & \multicolumn{1}{c|}{1.85}         & 0.67         \\ \cline{7-9} 
    \multicolumn{1}{|c|}{}                                    & \multicolumn{1}{c|}{}                      & \multicolumn{1}{c|}{}                      & \multicolumn{1}{c|}{}                                     & \multicolumn{1}{c|}{}                      & \multicolumn{1}{c|}{}                      & \multicolumn{1}{c|}{Both}                                & \multicolumn{1}{c|}{1.55}         & 0.75         \\ \cline{4-9} 
    \multicolumn{1}{|c|}{}                                    & \multicolumn{1}{c|}{}                      & \multicolumn{1}{c|}{}                      & \multicolumn{1}{c|}{\multirow{3}{*}{Neutral}}             & \multicolumn{1}{c|}{\multirow{3}{*}{2.11}} & \multicolumn{1}{c|}{\multirow{3}{*}{0.69}} & \multicolumn{1}{c|}{Male}                                & \multicolumn{1}{c|}{1.90}         & 0.85         \\ \cline{7-9} 
    \multicolumn{1}{|c|}{}                                    & \multicolumn{1}{c|}{}                      & \multicolumn{1}{c|}{}                      & \multicolumn{1}{c|}{}                                     & \multicolumn{1}{c|}{}                      & \multicolumn{1}{c|}{}                      & \multicolumn{1}{c|}{Female}                              & \multicolumn{1}{c|}{2.25}         & 0.55         \\ \cline{7-9} 
    \multicolumn{1}{|c|}{}                                    & \multicolumn{1}{c|}{}                      & \multicolumn{1}{c|}{}                      & \multicolumn{1}{c|}{}                                     & \multicolumn{1}{c|}{}                      & \multicolumn{1}{c|}{}                      & \multicolumn{1}{c|}{Both}                                & \multicolumn{1}{c|}{2.20}         & 0.61         \\ \cline{4-9} 
    \multicolumn{1}{|c|}{}                                    & \multicolumn{1}{c|}{}                      & \multicolumn{1}{c|}{}                      & \multicolumn{1}{c|}{\multirow{3}{*}{Positive}}            & \multicolumn{1}{c|}{\multirow{3}{*}{2.30}} & \multicolumn{1}{c|}{\multirow{3}{*}{0.84}} & \multicolumn{1}{c|}{Male}                                & \multicolumn{1}{c|}{2.15}         & 0.98         \\ \cline{7-9} 
    \multicolumn{1}{|c|}{}                                    & \multicolumn{1}{c|}{}                      & \multicolumn{1}{c|}{}                      & \multicolumn{1}{c|}{}                                     & \multicolumn{1}{c|}{}                      & \multicolumn{1}{c|}{}                      & \multicolumn{1}{c|}{Female}                              & \multicolumn{1}{c|}{2.50}         & 0.76         \\ \cline{7-9} 
    \multicolumn{1}{|c|}{}                                    & \multicolumn{1}{c|}{}                      & \multicolumn{1}{c|}{}                      & \multicolumn{1}{c|}{}                                     & \multicolumn{1}{c|}{}                      & \multicolumn{1}{c|}{}                      & \multicolumn{1}{c|}{Both}                                & \multicolumn{1}{c|}{2.25}         & 0.78         \\ \hline
    \end{tabular}
    \caption{Average and standard deviation of the User`s Feeling answers (Q1 in the questionnaire following Table~\ref{tab:questions}). For $Q1$ answer options, we attributed the following values to the next answers: 1 to negative, 2 to neutral, 3 to positive, and finally, 0 to I don't know.}
    \label{tab:user-feeling}
\end{table*}

Still in relation to $H0_2$, 
we found one significant result, using Question $Q4$, in relation to the main effect of VH Emotion (F(2, 537) = $19.76$ and $p<.001$; No equal variance (Stats = $12.72$ and $p=.001$)). With this, after the post-hoc test, we found significant results comparing the negative emotion and the other two conditions (neutral - $p<.001$; positive = $p=.024$), and between the neutral and positive emotions ($p<.001$). Table~\ref{tab:comfort} shows that the three highest average values related to the VH gender conditions are associated with the neutral emotion, and the three lowest with the negative emotion. 
Therefore, \textbf{we can say that people felt more comfortable with neutral emotion than the positive emotion (respectively, general averages of $3.20$ and $2.83$), more comfortable with neutral emotion than the negative emotion (respectively, general averages of $3.20$ and $2.57$), and more comfortable with positive emotion than the negative emotion (respectively, general averages of $2.83$ and $2.57$).} In general, \textbf{we can reject partially $H0_2$, since people felt emotionally and comfortable differently towards groups of VHs with different emotions, and felt emotionally and comfortable similarly towards groups of VHs with different genders.}


\begin{table*}[!htb]
    \centering
    \begin{tabular}{|ccccccccc|}
    \hline
    \multicolumn{9}{|c|}{\textbf{Comfort Answers}}                                                                                                                                                                                                                                                                                                                                                                                   \\ \hline
    \multicolumn{1}{|c|}{\multirow{2}{*}{\textbf{Participant Gender}}} & \multicolumn{2}{c|}{\textbf{User's Answer}}                                             & \multicolumn{1}{c|}{\multirow{2}{*}{\textbf{VH Emotion}}} & \multicolumn{2}{c|}{\textbf{User's Answer}}                                             & \multicolumn{1}{c|}{\multirow{2}{*}{\textbf{VH Gender}}} & \multicolumn{2}{c|}{\textbf{User's Answer}}      \\ \cline{2-3} \cline{5-6} \cline{8-9} 
    \multicolumn{1}{|c|}{}                                             & \multicolumn{1}{c|}{\textbf{Avg}}          & \multicolumn{1}{c|}{\textbf{STD}}          & \multicolumn{1}{c|}{}                                     & \multicolumn{1}{c|}{\textbf{Avg}}          & \multicolumn{1}{c|}{\textbf{STD}}          & \multicolumn{1}{c|}{}                                    & \multicolumn{1}{c|}{\textbf{Avg}} & \textbf{STD} \\ \hline
    \multicolumn{1}{|c|}{\multirow{9}{*}{Male}}                        & \multicolumn{1}{c|}{\multirow{9}{*}{2.82}} & \multicolumn{1}{c|}{\multirow{9}{*}{0.87}} & \multicolumn{1}{c|}{\multirow{3}{*}{Negative}}            & \multicolumn{1}{c|}{\multirow{3}{*}{2.52}} & \multicolumn{1}{c|}{\multirow{3}{*}{0.78}} & \multicolumn{1}{c|}{Male}                                & \multicolumn{1}{c|}{2.60}         & 0.77         \\ \cline{7-9} 
    \multicolumn{1}{|c|}{}                                             & \multicolumn{1}{c|}{}                      & \multicolumn{1}{c|}{}                      & \multicolumn{1}{c|}{}                                     & \multicolumn{1}{c|}{}                      & \multicolumn{1}{c|}{}                      & \multicolumn{1}{c|}{Female}                              & \multicolumn{1}{c|}{2.27}         & 0.78         \\ \cline{7-9} 
    \multicolumn{1}{|c|}{}                                             & \multicolumn{1}{c|}{}                      & \multicolumn{1}{c|}{}                      & \multicolumn{1}{c|}{}                                     & \multicolumn{1}{c|}{}                      & \multicolumn{1}{c|}{}                      & \multicolumn{1}{c|}{Both}                                & \multicolumn{1}{c|}{2.70}         & 0.75         \\ \cline{4-9} 
    \multicolumn{1}{|c|}{}                                             & \multicolumn{1}{c|}{}                      & \multicolumn{1}{c|}{}                      & \multicolumn{1}{c|}{\multirow{3}{*}{Neutral}}             & \multicolumn{1}{c|}{\multirow{3}{*}{3.19}} & \multicolumn{1}{c|}{\multirow{3}{*}{0.73}} & \multicolumn{1}{c|}{Male}                                & \multicolumn{1}{c|}{3.17}         & 0.67         \\ \cline{7-9} 
    \multicolumn{1}{|c|}{}                                             & \multicolumn{1}{c|}{}                      & \multicolumn{1}{c|}{}                      & \multicolumn{1}{c|}{}                                     & \multicolumn{1}{c|}{}                      & \multicolumn{1}{c|}{}                      & \multicolumn{1}{c|}{Female}                              & \multicolumn{1}{c|}{3.20}         & 0.75         \\ \cline{7-9} 
    \multicolumn{1}{|c|}{}                                             & \multicolumn{1}{c|}{}                      & \multicolumn{1}{c|}{}                      & \multicolumn{1}{c|}{}                                     & \multicolumn{1}{c|}{}                      & \multicolumn{1}{c|}{}                      & \multicolumn{1}{c|}{Both}                                & \multicolumn{1}{c|}{3.20}         & 0.79         \\ \cline{4-9} 
    \multicolumn{1}{|c|}{}                                             & \multicolumn{1}{c|}{}                      & \multicolumn{1}{c|}{}                      & \multicolumn{1}{c|}{\multirow{3}{*}{Positive}}            & \multicolumn{1}{c|}{\multirow{3}{*}{2.76}} & \multicolumn{1}{c|}{\multirow{3}{*}{0.96}} & \multicolumn{1}{c|}{Male}                                & \multicolumn{1}{c|}{2.70}         & 0.96         \\ \cline{7-9} 
    \multicolumn{1}{|c|}{}                                             & \multicolumn{1}{c|}{}                      & \multicolumn{1}{c|}{}                      & \multicolumn{1}{c|}{}                                     & \multicolumn{1}{c|}{}                      & \multicolumn{1}{c|}{}                      & \multicolumn{1}{c|}{Female}                              & \multicolumn{1}{c|}{2.92}         & 0.88         \\ \cline{7-9} 
    \multicolumn{1}{|c|}{}                                             & \multicolumn{1}{c|}{}                      & \multicolumn{1}{c|}{}                      & \multicolumn{1}{c|}{}                                     & \multicolumn{1}{c|}{}                      & \multicolumn{1}{c|}{}                      & \multicolumn{1}{c|}{Both}                                & \multicolumn{1}{c|}{2.67}         & 1.04         \\ \hline
    \multicolumn{1}{|c|}{\multirow{9}{*}{Female}}                      & \multicolumn{1}{c|}{\multirow{9}{*}{2.96}} & \multicolumn{1}{c|}{\multirow{9}{*}{0.94}} & \multicolumn{1}{c|}{\multirow{3}{*}{Negative}}            & \multicolumn{1}{c|}{\multirow{3}{*}{2.68}} & \multicolumn{1}{c|}{\multirow{3}{*}{0.87}} & \multicolumn{1}{c|}{Male}                                & \multicolumn{1}{c|}{2.70}         & 0.92         \\ \cline{7-9} 
    \multicolumn{1}{|c|}{}                                             & \multicolumn{1}{c|}{}                      & \multicolumn{1}{c|}{}                      & \multicolumn{1}{c|}{}                                     & \multicolumn{1}{c|}{}                      & \multicolumn{1}{c|}{}                      & \multicolumn{1}{c|}{Female}                              & \multicolumn{1}{c|}{2.75}         & 0.91         \\ \cline{7-9} 
    \multicolumn{1}{|c|}{}                                             & \multicolumn{1}{c|}{}                      & \multicolumn{1}{c|}{}                      & \multicolumn{1}{c|}{}                                     & \multicolumn{1}{c|}{}                      & \multicolumn{1}{c|}{}                      & \multicolumn{1}{c|}{Both}                                & \multicolumn{1}{c|}{2.60}         & 0.82         \\ \cline{4-9} 
    \multicolumn{1}{|c|}{}                                             & \multicolumn{1}{c|}{}                      & \multicolumn{1}{c|}{}                      & \multicolumn{1}{c|}{\multirow{3}{*}{Neutral}}             & \multicolumn{1}{c|}{\multirow{3}{*}{3.23}} & \multicolumn{1}{c|}{\multirow{3}{*}{0.85}} & \multicolumn{1}{c|}{Male}                                & \multicolumn{1}{c|}{3.10}         & 0.91         \\ \cline{7-9} 
    \multicolumn{1}{|c|}{}                                             & \multicolumn{1}{c|}{}                      & \multicolumn{1}{c|}{}                      & \multicolumn{1}{c|}{}                                     & \multicolumn{1}{c|}{}                      & \multicolumn{1}{c|}{}                      & \multicolumn{1}{c|}{Female}                              & \multicolumn{1}{c|}{3.40}         & 0.82         \\ \cline{7-9} 
    \multicolumn{1}{|c|}{}                                             & \multicolumn{1}{c|}{}                      & \multicolumn{1}{c|}{}                      & \multicolumn{1}{c|}{}                                     & \multicolumn{1}{c|}{}                      & \multicolumn{1}{c|}{}                      & \multicolumn{1}{c|}{Both}                                & \multicolumn{1}{c|}{3.20}         & 0.83         \\ \cline{4-9} 
    \multicolumn{1}{|c|}{}                                             & \multicolumn{1}{c|}{}                      & \multicolumn{1}{c|}{}                      & \multicolumn{1}{c|}{\multirow{3}{*}{Positive}}            & \multicolumn{1}{c|}{\multirow{3}{*}{2.29}} & \multicolumn{1}{c|}{\multirow{3}{*}{1.04}} & \multicolumn{1}{c|}{Male}                                & \multicolumn{1}{c|}{3.00}         & 1.12         \\ \cline{7-9} 
    \multicolumn{1}{|c|}{}                                             & \multicolumn{1}{c|}{}                      & \multicolumn{1}{c|}{}                      & \multicolumn{1}{c|}{}                                     & \multicolumn{1}{c|}{}                      & \multicolumn{1}{c|}{}                      & \multicolumn{1}{c|}{Female}                              & \multicolumn{1}{c|}{3.15}         & 0.93         \\ \cline{7-9} 
    \multicolumn{1}{|c|}{}                                             & \multicolumn{1}{c|}{}                      & \multicolumn{1}{c|}{}                      & \multicolumn{1}{c|}{}                                     & \multicolumn{1}{c|}{}                      & \multicolumn{1}{c|}{}                      & \multicolumn{1}{c|}{Both}                                & \multicolumn{1}{c|}{2.75}         & 1.06         \\ \hline
    \end{tabular}
    \caption{Average and standard deviation of the participant Comfort answers (Q4 in the questionnaire following Table~\ref{tab:questions}). For $Q5$ answer options, we followed the 5-Likert scale where very uncomfortable is 1 and very comfortable is 5.}
    \label{tab:comfort}
\end{table*}

\subsection{Analysis of Realism Perception}
\label{sec:realism}

For statistics analysis of $H0_3$, we focused on $Q5$, as presented in Table~\ref{tab:questions}. For $Q5$, to have a realism score, we followed the 5-Likert scale where very unrealistic is 1 and very realistic is 5.

Considering the $H0_3$, we found two significant results in the statistical test using the Question $Q5$. 
The first was the main effect of VH Gender (F(2, 537) = $3.68$ and $p=.025$), and the other was the main effect of VH Emotion (F(2, 537) = $11.70$ and $p<.001$). In relation to the VH Gender effect, we performed the post-hoc test after observing that the data have equal variance (Stats = $3.00$ and $p=.222$). We only found a significant result between female VH and male VH ($p=.009$). As we can see from the realism scores in Table~\ref{tab:realism}, the groups that had only female VHs had higher realism averages than when they had only male VHs.
Therefore, \textbf{we can say that people perceived that the female VH is more realistic than male VH (respectively, general averages of $3.28$ and $3.01$). 
} Analysing the main effect of the VH Emotion, the data did not have equal variance (Stats = $8.05$ and $p=.017$). We found significant results comparing the negative emotion with the neutral emotion ($p<.001$) and the neutral emotion with the positive emotion ($p<.001$). Therefore, \textbf{we can say people perceived more realism in neutral emotion than in negative emotion (respectively, general averages of $3.41$ and $3.03$), more realism in neutral emotion than in positive emotion (respectively, general averages of $3.41$ and $2.92$), and people perceived similar realism between positive emotion and negative emotion.} In this case, \textbf{we can reject $H0_3$ in relation to the emotion and VH genders, as people perceived realism differently in each group of VHs with different emotions, and perceived more realism in the group that contained only female VHs than in the group that only had male VHs.}

\begin{table*}[!htb]
    \centering
    \begin{tabular}{|ccccccccc|}
    \hline
    \multicolumn{9}{|c|}{\textbf{Realism Answers}}                                                                                                                                                                                                                                                                                                                                                                                   \\ \hline
    \multicolumn{1}{|c|}{\multirow{2}{*}{\textbf{Participant Gender}}} & \multicolumn{2}{c|}{\textbf{User's Answer}}                                             & \multicolumn{1}{c|}{\multirow{2}{*}{\textbf{VH Emotion}}} & \multicolumn{2}{c|}{\textbf{User's Answer}}                                             & \multicolumn{1}{c|}{\multirow{2}{*}{\textbf{VH Gender}}} & \multicolumn{2}{c|}{\textbf{User's Answer}}      \\ \cline{2-3} \cline{5-6} \cline{8-9} 
    \multicolumn{1}{|c|}{}                                             & \multicolumn{1}{c|}{\textbf{Avg}}          & \multicolumn{1}{c|}{\textbf{STD}}          & \multicolumn{1}{c|}{}                                     & \multicolumn{1}{c|}{\textbf{Avg}}          & \multicolumn{1}{c|}{\textbf{STD}}          & \multicolumn{1}{c|}{}                                    & \multicolumn{1}{c|}{\textbf{Avg}} & \textbf{STD} \\ \hline
    \multicolumn{1}{|c|}{\multirow{9}{*}{Male}}                        & \multicolumn{1}{c|}{\multirow{9}{*}{3.10}} & \multicolumn{1}{c|}{\multirow{9}{*}{0.86}} & \multicolumn{1}{c|}{\multirow{3}{*}{Negative}}            & \multicolumn{1}{c|}{\multirow{3}{*}{3.01}} & \multicolumn{1}{c|}{\multirow{3}{*}{0.49}} & \multicolumn{1}{c|}{Male}                                & \multicolumn{1}{c|}{2.80}         & 0.88         \\ \cline{7-9} 
    \multicolumn{1}{|c|}{}                                             & \multicolumn{1}{c|}{}                      & \multicolumn{1}{c|}{}                      & \multicolumn{1}{c|}{}                                     & \multicolumn{1}{c|}{}                      & \multicolumn{1}{c|}{}                      & \multicolumn{1}{c|}{Female}                              & \multicolumn{1}{c|}{3.27}         & 0.75         \\ \cline{7-9} 
    \multicolumn{1}{|c|}{}                                             & \multicolumn{1}{c|}{}                      & \multicolumn{1}{c|}{}                      & \multicolumn{1}{c|}{}                                     & \multicolumn{1}{c|}{}                      & \multicolumn{1}{c|}{}                      & \multicolumn{1}{c|}{Both}                                & \multicolumn{1}{c|}{2.97}         & 0.69         \\ \cline{4-9} 
    \multicolumn{1}{|c|}{}                                             & \multicolumn{1}{c|}{}                      & \multicolumn{1}{c|}{}                      & \multicolumn{1}{c|}{\multirow{3}{*}{Neutral}}             & \multicolumn{1}{c|}{\multirow{3}{*}{3.45}} & \multicolumn{1}{c|}{\multirow{3}{*}{0.73}} & \multicolumn{1}{c|}{Male}                                & \multicolumn{1}{c|}{3.32}         & 0.82         \\ \cline{7-9} 
    \multicolumn{1}{|c|}{}                                             & \multicolumn{1}{c|}{}                      & \multicolumn{1}{c|}{}                      & \multicolumn{1}{c|}{}                                     & \multicolumn{1}{c|}{}                      & \multicolumn{1}{c|}{}                      & \multicolumn{1}{c|}{Female}                              & \multicolumn{1}{c|}{3.57}         & 0.59         \\ \cline{7-9} 
    \multicolumn{1}{|c|}{}                                             & \multicolumn{1}{c|}{}                      & \multicolumn{1}{c|}{}                      & \multicolumn{1}{c|}{}                                     & \multicolumn{1}{c|}{}                      & \multicolumn{1}{c|}{}                      & \multicolumn{1}{c|}{Both}                                & \multicolumn{1}{c|}{3.45}         & 0.74         \\ \cline{4-9} 
    \multicolumn{1}{|c|}{}                                             & \multicolumn{1}{c|}{}                      & \multicolumn{1}{c|}{}                      & \multicolumn{1}{c|}{\multirow{3}{*}{Positive}}            & \multicolumn{1}{c|}{\multirow{3}{*}{2.85}} & \multicolumn{1}{c|}{\multirow{3}{*}{0.95}} & \multicolumn{1}{c|}{Male}                                & \multicolumn{1}{c|}{2.80}         & 0.93         \\ \cline{7-9} 
    \multicolumn{1}{|c|}{}                                             & \multicolumn{1}{c|}{}                      & \multicolumn{1}{c|}{}                      & \multicolumn{1}{c|}{}                                     & \multicolumn{1}{c|}{}                      & \multicolumn{1}{c|}{}                      & \multicolumn{1}{c|}{Female}                              & \multicolumn{1}{c|}{3.05}         & 0.87         \\ \cline{7-9} 
    \multicolumn{1}{|c|}{}                                             & \multicolumn{1}{c|}{}                      & \multicolumn{1}{c|}{}                      & \multicolumn{1}{c|}{}                                     & \multicolumn{1}{c|}{}                      & \multicolumn{1}{c|}{}                      & \multicolumn{1}{c|}{Both}                                & \multicolumn{1}{c|}{2.72}         & 1.03         \\ \hline
    \multicolumn{1}{|c|}{\multirow{9}{*}{Female}}                      & \multicolumn{1}{c|}{\multirow{9}{*}{3.16}} & \multicolumn{1}{c|}{\multirow{9}{*}{0.92}} & \multicolumn{1}{c|}{\multirow{3}{*}{Negative}}            & \multicolumn{1}{c|}{\multirow{3}{*}{3.06}} & \multicolumn{1}{c|}{\multirow{3}{*}{0.93}} & \multicolumn{1}{c|}{Male}                                & \multicolumn{1}{c|}{3.00}         & 0.97         \\ \cline{7-9} 
    \multicolumn{1}{|c|}{}                                             & \multicolumn{1}{c|}{}                      & \multicolumn{1}{c|}{}                      & \multicolumn{1}{c|}{}                                     & \multicolumn{1}{c|}{}                      & \multicolumn{1}{c|}{}                      & \multicolumn{1}{c|}{Female}                              & \multicolumn{1}{c|}{3.15}         & 0.93         \\ \cline{7-9} 
    \multicolumn{1}{|c|}{}                                             & \multicolumn{1}{c|}{}                      & \multicolumn{1}{c|}{}                      & \multicolumn{1}{c|}{}                                     & \multicolumn{1}{c|}{}                      & \multicolumn{1}{c|}{}                      & \multicolumn{1}{c|}{Both}                                & \multicolumn{1}{c|}{3.05}         & 0.94         \\ \cline{4-9} 
    \multicolumn{1}{|c|}{}                                             & \multicolumn{1}{c|}{}                      & \multicolumn{1}{c|}{}                      & \multicolumn{1}{c|}{\multirow{3}{*}{Neutral}}             & \multicolumn{1}{c|}{\multirow{3}{*}{3.35}} & \multicolumn{1}{c|}{\multirow{3}{*}{0.86}} & \multicolumn{1}{c|}{Male}                                & \multicolumn{1}{c|}{3.20}         & 0.89         \\ \cline{7-9} 
    \multicolumn{1}{|c|}{}                                             & \multicolumn{1}{c|}{}                      & \multicolumn{1}{c|}{}                      & \multicolumn{1}{c|}{}                                     & \multicolumn{1}{c|}{}                      & \multicolumn{1}{c|}{}                      & \multicolumn{1}{c|}{Female}                              & \multicolumn{1}{c|}{3.50}         & 0.82         \\ \cline{7-9} 
    \multicolumn{1}{|c|}{}                                             & \multicolumn{1}{c|}{}                      & \multicolumn{1}{c|}{}                      & \multicolumn{1}{c|}{}                                     & \multicolumn{1}{c|}{}                      & \multicolumn{1}{c|}{}                      & \multicolumn{1}{c|}{Both}                                & \multicolumn{1}{c|}{3.35}         & 0.87         \\ \cline{4-9} 
    \multicolumn{1}{|c|}{}                                             & \multicolumn{1}{c|}{}                      & \multicolumn{1}{c|}{}                      & \multicolumn{1}{c|}{\multirow{3}{*}{Positive}}            & \multicolumn{1}{c|}{\multirow{3}{*}{3.06}} & \multicolumn{1}{c|}{\multirow{3}{*}{0.95}} & \multicolumn{1}{c|}{Male}                                & \multicolumn{1}{c|}{3.10}         & 1.02         \\ \cline{7-9} 
    \multicolumn{1}{|c|}{}                                             & \multicolumn{1}{c|}{}                      & \multicolumn{1}{c|}{}                      & \multicolumn{1}{c|}{}                                     & \multicolumn{1}{c|}{}                      & \multicolumn{1}{c|}{}                      & \multicolumn{1}{c|}{Female}                              & \multicolumn{1}{c|}{3.15}         & 0.93         \\ \cline{7-9} 
    \multicolumn{1}{|c|}{}                                             & \multicolumn{1}{c|}{}                      & \multicolumn{1}{c|}{}                      & \multicolumn{1}{c|}{}                                     & \multicolumn{1}{c|}{}                      & \multicolumn{1}{c|}{}                      & \multicolumn{1}{c|}{Both}                                & \multicolumn{1}{c|}{2.95}         & 0.94         \\ \hline
    \end{tabular}
    \caption{Average and standard deviation of the realism answers (Q5 in the questionnaire following Table~\ref{tab:questions}). For $Q5$ answer options, we followed the 5-Likert scale where very unrealistic is 1 and very realistic is 5}
    \label{tab:realism}
\end{table*}

\section{Discussion}
This section aims to discuss the results regarding the hypotheses of our work. Overall, we found no effect of participant gender. This means that in our research, there were no in-group effects of gender, i.e., the gender of the VHs did not influence the responses of people with the same gender as them. 
In relation to emotion recognition in general, our findings indicate that the emotions expressed by VHs were perceived accurately by people, independent on the VH gender 
and the emotion itself. However, when we only looked at negative emotions, the gender of the VHs influenced emotion recognition, where people were more accurate in recognizing the negative emotion in the group of only female VHs than in the group of only men. 

Regarding emotions, people were correct in recognizing the happy expression as a positive emotion, the neutral expression as a neutral emotion, and the anger expression as a negative emotion. In this case, this result also shows the accuracy of our methodology for modeling the emotions of VHs, although using simple characters animated directly from performance driven animation tools. Positive emotion, in particular, was distinctly considered more intense compared to neutral and negative emotions. This may reflect an innate human tendency to identify more strongly with the emotion of happiness and is in accordance with literature~\cite{elfenbein2007toward}. Furthermore, the interaction between VH gender and emotion revealed that negative emotion was perceived more intensely in female VHs compared to male VHs. This suggests that gender expectations might play a role in how we interpret emotions in virtual contexts. This result goes in the same direction as the result of the work by McDuff et al.~\cite{mcduff2017large}, which showed that women tend to be considered more expressive (positively and negatively) than men. And also, in part, in the same line as Durupinar and Kim's work~\cite{durupinar2022facial}, where the results showed that female VHs had more easily recognized emotions than male VHs. However, as mentioned in the work of Zibrek et al.~\cite{zibrek2015exploring}, anger (negative emotion used in our work) is an emotion more attributed to men. In fact, the results of the work by Zibrek and colleagues showed that people recognized the emotion anger more in male VHs than in female VHs, that is, contrary to our result. In this case, what may have happened in the present work is that the groups of VHs may have amplified the emotional recognition~\cite{goldenberg2021crowd}. For example, in some psychology studies, the emotional recognition in groups is different from the emotional recognition in individuals~\cite{goldenberg2021crowd,lamer2018rapid,whitney2018ensemble}, including in groups with different genders~\cite{haberman2007rapid}.

In relation to hypothesis $H0_2$, participants reported varying levels of comfort in response to the VHs' emotions. Positive and neutral emotions were associated with greater comfort, while negative emotions were linked to lesser comfort. Regarding positive emotion, which in this case was the happy emotion, the result of perceived comfort was similar to the work of Tinwell et al.~\cite{tinwell2011facial}, which showed that people did not feel negative emotional valences (discomfort, uncanny, strangeness, etc) in relation to VH expressing happiness. However, in the case of the emotion anger (negative), in the work of Tinwell and colleagues, this emotion also did not convey feelings of strangeness, which goes against our results. In this case, our result on anger is in the same line as the result presented in the work of M{\"a}k{\"a}r{\"a}inen et al.~\cite{makarainen2014exaggerating}, the emotion of anger did not reduce the feeling of strangeness. In our case, the feeling of strangeness may result from the simple technology used rather than from other characteristics such as gender and grouping. Future work here is desirable to better analyse the results.

Our results refuted $H0_3$, as the perception of realism in VHs differed based on the gender and emotion of the VH. Female VHs were perceived as more realistic than male VHs. This result is in line with the work of Araujo et al.~\cite{araujo2021analysis}, where people perceived more realism in female VHs than male ones. In terms of emotions, neutral emotion was considered more realistic than negative and positive emotions. In this case, considering that the muscles on the VHs' faces remained static during the neutral emotion, and moved during positive and negative emotions, we can say that facial movement influenced the perception of realism. Looking at the results of hypotheses $H0_2$ and $H0_3$, we can make an analogy with UV~\cite{mori2012uncanny} and its movement hypothesis~\cite{mori2012uncanny,katsyri2015review}, which states that at certain points of human likeness levels (perceived realism), artificial beings in static stimuli can convey more comfort than when they are in motion. In our case, the group of VHs with neutral faces (without facial movements) conveyed more comfort and more realism than the group with negative emotion, and only more realism than the groups with positive emotion, that is, with facial movements. However, as the UV movement hypothesis did not take into account groups of artificial beings, more analyzes based on groups of VHs with different levels of realism are needed.


In general, still looking at the results involving perceived comfort and realism, which always involved cartoonish VHs, and comparing with the literature involving photorealistic VHs, we could see similar results. For example, in the work of Higgs et al.~\cite{higgins2023investigating}, which used Metahumans and UV-based evaluation metrics, the results showed that positive emotions were considered more appealing than negative emotions. In fact, Higgs et al. even measured Metahumans with different levels of realism, and the results showed that the lower level of realism led to higher reports of happy emotion appeal compared to neutral and sad conditions. The same happened in our work regarding perceived comfort, where participants felt more comfortable with positive emotions than negative ones. 
In other words, our results may indicate that less realistic VHs are still able to convey more comfort in positive than negative emotions, just like very realistic VHs used as state of the art.


\section{Final Considerations}
\label{sec:final_considerations}

This work presents a perception study on how people feel and perceive comfort and realism in interactive and simple cartoon VHs. Once again, we explore available and affordable technologies to create VHs that populate a virtual environment. 
The summary of our results are: Firstly, we rejected $H0_1$ (People recognize emotions similarly in groups of VHs with different genders and emotions), indicating that people accurately recognized the different emotions expressed by groups of VHs. Additionally, 
our results revealed that the negative emotion of the female VH was recognized as more negative than the negative male VH. We partially rejected $H0_2$ (People experience similar comfort towards groups of VHs with different genders and emotions), demonstrating that people perceived comfort differently depending on the emotion
. In this case, we observed that participants felt better with positive and neutral emotions than with negative emotions but felt the most comfort with the neutral emotion. Furthermore, we could not completely reject the $H0_2$ as the gender of the VHs did not influence the participants' perceived comfort.
Finally, we rejected $H0_3$ (People perceive the realism of groups of VHs with different genders and emotions similarly), indicating that people perceived more realism in the female VH than in the male VH, and perceived more realism in the neutral emotion than in the others. 

We believe that studies on the perception of groups and crowds of VHs can be important for both the area of psychology and the area of computer graphics. In relation to psychology, our work can contribute to studies on in-group advantage, contexts of group interactions in social environments, differences between genders and emotions, etc. In relation to computer graphics, our work can contribute to interaction in games, more realistic simulations in relation to group dynamics, etc. 
For future work, we intend to test the same hypotheses with VHs having different levels of realism, groups with more VHs, greater gender diversity of VHs, diversity of race and skin color, and different environments.
%
%
%
\bibliographystyle{splncs04}
\bibliography{ref} 

\begin{thebibliography}{10}
\providecommand{\url}[1]{\texttt{#1}}
\providecommand{\urlprefix}{URL }
\providecommand{\doi}[1]{https://doi.org/#1}

\bibitem{abbruzzese2019age}
Abbruzzese, L., Magnani, N., Robertson, I.H., Mancuso, M.: Age and gender differences in emotion recognition. Frontiers in psychology  \textbf{10}, ~2371 (2019)

\bibitem{alkawaz2015blend}
Alkawaz, M.H., Mohamad, D., Basori, A.H., Saba, T.: Blend shape interpolation and facs for realistic avatar. 3D Research  \textbf{6},  1--10 (2015)

\bibitem{andreotti2021perception}
Andreotti, L., Weber, M.L., da~Silva, T.L., de~Andrade~Araujo, V.F., Musse, S.R.: Perception of charisma, comfort, micro and macro expressions in computer graphics characters. In: 2021 20th Brazilian Symposium on Computer Games and Digital Entertainment (SBGames). pp. 107--116. IEEE (2021)

\bibitem{araujo2021much}
Araujo, V., Dalmoro, B., Favaretto, R., Vilanova, F., Costa, A., Musse, S.R.: How much do we perceive geometric features, personalities and emotions in avatars? In: Advances in Computer Graphics: 38th Computer Graphics International Conference, CGI 2021, Virtual Event, September 6--10, 2021, Proceedings 38. pp. 548--567. Springer (2021)

\bibitem{araujo2021analysis}
Araujo, V., Dalmoro, B., Musse, S.R.: Analysis of charisma, comfort and realism in cg characters from a gender perspective. The Visual Computer  \textbf{37}(9-11),  2685--2698 (2021)

\bibitem{araujo2021perceived}
Araujo, V., Melgare, J., Dalmoro, B.M., Musse, S.R.: Is the perceived comfort with cg characters increasing with their novelty? IEEE Computer Graphics and Applications  \textbf{42}(1),  32--46 (2021)

\bibitem{bailey2017gender}
Bailey, J.D., Blackmore, K.L.: Gender and the perception of emotions in avatars. In: Proceedings of the Australasian Computer Science Week Multiconference. pp.~1--8 (2017)

\bibitem{bassili1978facial}
Bassili, J.N.: Facial motion in the perception of faces and of emotional expression. Journal of experimental psychology: human perception and performance  \textbf{4}(3), ~373 (1978)

\bibitem{brown2020social}
Brown, R.: The social identity approach: Appraising the tajfellian legacy. British Journal of Social Psychology  \textbf{59}(1),  5--25 (2020)

\bibitem{campbell1958common}
Campbell, D.T.: Common fate, similarity, and other indices of the status of aggregates of persons as social entities. Behavioral science  \textbf{3}(1), ~14 (1958)

\bibitem{cohn2009happiness}
Cohn, M.A., Fredrickson, B.L., Brown, S.L., Mikels, J.A., Conway, A.M.: Happiness unpacked: positive emotions increase life satisfaction by building resilience. Emotion  \textbf{9}(3), ~361 (2009)

\bibitem{draude2011intermediaries}
Draude, C.: Intermediaries: reflections on virtual humans, gender, and the uncanny valley. AI \& society  \textbf{26},  319--327 (2011)

\bibitem{durupinar2022facial}
Durupinar, F., Kim, J.: Facial emotion recognition of virtual humans with different genders, races, and ages. In: ACM Symposium on Applied Perception 2022. pp. 1--10 (2022)

\bibitem{Ekman2013AnAF}
Ekman, P.: An argument for basic emotions (2013)

\bibitem{ekman1978facial}
Ekman, P., Friesen, W.V.: Facial action coding system. Environmental Psychology \& Nonverbal Behavior  (1978)

\bibitem{EkmanFACS}
Ekman, P., V.~Friesen, W.: Facial action coding system (facs)  (01 2002)

\bibitem{elfenbein2002there}
Elfenbein, H.A., Ambady, N.: Is there an in-group advantage in emotion recognition?  (2002)

\bibitem{elfenbein2007toward}
Elfenbein, H.A., Beaupr{\'e}, M., L{\'e}vesque, M., Hess, U.: Toward a dialect theory: cultural differences in the expression and recognition of posed facial expressions. Emotion  \textbf{7}(1), ~131 (2007)

\bibitem{elias2017ensemble}
Elias, E., Dyer, M., Sweeny, T.D.: Ensemble perception of dynamic emotional groups. Psychological Science  \textbf{28}(2),  193--203 (2017)

\bibitem{ennis2013emotion}
Ennis, C., Hoyet, L., Egges, A., McDonnell, R.: Emotion capture: Emotionally expressive characters for games. In: Proceedings of motion on games, pp. 53--60 (2013)

\bibitem{Favarettl2019}
Favaretto, R.M., Knob, P., Musse, S.R., Vilanova, F., Costa, A.B.: Detecting personality and emotion traits in crowds from video sequences. Mach. Vision Appl.  \textbf{30}(5),  999–1012 (Jul 2019). \doi{10.1007/s00138-018-0979-y}, \url{https://doi.org/10.1007/s00138-018-0979-y}

\bibitem{friesen1983emfacs}
Friesen, W.V., Ekman, P., et~al.: Emfacs-7: Emotional facial action coding system. Unpublished manuscript, University of California at San Francisco  \textbf{2}(36), ~1 (1983)

\bibitem{ghosh4evaluating}
Ghosh, R., Feijoo-Garcia, P.G., Wrenn, C., Stuart, J., Lok, B.: Evaluating face gender cues in virtual humans within and beyond the gender binary. Frontiers in Virtual Reality  \textbf{4},  1251420 (2023)

\bibitem{goldenberg2021crowd}
Goldenberg, A., Weisz, E., Sweeny, T.D., Cikara, M., Gross, J.J.: The crowd-emotion-amplification effect. Psychological science  \textbf{32}(3),  437--450 (2021)

\bibitem{guadagno2007virtual}
Guadagno, R.E., Blascovich, J., Bailenson, J.N., McCall, C.: Virtual humans and persuasion: The effects of agency and behavioral realism. Media Psychology  \textbf{10}(1),  1--22 (2007)

\bibitem{haberman2007rapid}
Haberman, J., Whitney, D.: Rapid extraction of mean emotion and gender from sets of faces. Current biology  \textbf{17}(17),  R751--R753 (2007)

\bibitem{higgins2023investigating}
Higgins, D., Zhan, Y., Cowan, B.R., McDonnell, R.: Investigating the effect of visual realism on empathic responses to emotionally expressive virtual humans. In: ACM Symposium on Applied Perception 2023. pp.~1--7 (2023)

\bibitem{johansson1973visual}
Johansson, G.: Visual perception of biological motion and a model for its analysis. Perception \& psychophysics  \textbf{14}(2),  201--211 (1973)

\bibitem{joshi2006learning}
Joshi, P., Tien, W.C., Desbrun, M., Pighin, F.: Learning controls for blend shape based realistic facial animation. In: ACM Siggraph 2006 Courses, pp. 17--es (2006)

\bibitem{katsyri2015review}
K{\"a}tsyri, J., F{\"o}rger, K., M{\"a}k{\"a}r{\"a}inen, M., Takala, T.: A review of empirical evidence on different uncanny valley hypotheses: support for perceptual mismatch as one road to the valley of eeriness. Frontiers in psychology  \textbf{6}, ~390 (2015)

\bibitem{kendon1990conducting}
Kendon, A.: Conducting interaction: Patterns of behavior in focused encounters, vol.~7. CUP Archive (1990)

\bibitem{krumhuber2015real}
Krumhuber, E.G., Swiderska, A., Tsankova, E., Kamble, S.V., Kappas, A.: Real or artificial? intergroup biases in mind perception in a cross-cultural perspective. PLoS One  \textbf{10}(9),  e0137840 (2015)

\bibitem{lamer2018rapid}
Lamer, S.A., Sweeny, T.D., Dyer, M.L., Weisbuch, M.: Rapid visual perception of interracial crowds: Racial category learning from emotional segregation. Journal of Experimental Psychology: General  \textbf{147}(5), ~683 (2018)

\bibitem{macdorman2016reducing}
MacDorman, K.F., Chattopadhyay, D.: Reducing consistency in human realism increases the uncanny valley effect; increasing category uncertainty does not. Cognition  \textbf{146},  190--205 (2016)

\bibitem{makarainen2014exaggerating}
M{\"a}k{\"a}r{\"a}inen, M., K{\"a}tsyri, J., Takala, T.: Exaggerating facial expressions: A way to intensify emotion or a way to the uncanny valley? Cognitive Computation  \textbf{6},  708--721 (2014)

\bibitem{Tharindu2019}
Mathew, C.D.T., Knob, P.R., Musse, S.R., Aliaga, D.G.: Urban walkability design using virtual population simulation. Computer Graphics Forum  \textbf{38}(1),  455--469 (2019). \doi{https://doi.org/10.1111/cgf.13585}, \url{https://onlinelibrary.wiley.com/doi/abs/10.1111/cgf.13585}

\bibitem{mcdonnell2007virtual}
McDonnell, R., J{\"o}rg, S., Hodgins, J.K., Newell, F., O'Sullivan, C.: Virtual shapers \& movers: form and motion affect sex perception. In: Proceedings of the 4th symposium on Applied perception in graphics and visualization. pp. 7--10 (2007)

\bibitem{mcdonnell2008clone}
McDonnell, R., Larkin, M., Dobbyn, S., Collins, S., O'Sullivan, C.: Clone attack! perception of crowd variety. In: ACM SIGGRAPH 2008 papers, pp.~1--8 (2008)

\bibitem{mcduff2017large}
McDuff, D., Kodra, E., Kaliouby, R.e., LaFrance, M.: A large-scale analysis of sex differences in facial expressions. PloS one  \textbf{12}(4),  e0173942 (2017)

\bibitem{mori2012uncanny}
Mori, M., MacDorman, K.F., Kageki, N.: The uncanny valley [from the field]. IEEE Robotics \& automation magazine  \textbf{19}(2),  98--100 (2012)

\bibitem{musse2021history}
Musse, S.R., Cassol, V.J., Thalmann, D.: A history of crowd simulation: the past, evolution, and new perspectives. The Visual Computer pp. 1--16 (2021)

\bibitem{queiroz2014investigating}
Queiroz, R.B., Musse, S.R., Badler, N.I.: Investigating macroexpressions and microexpressions in computer graphics animated faces. Presence  \textbf{23}(2),  191--208 (2014)

\bibitem{tinwell2011facial}
Tinwell, A., Grimshaw, M., Nabi, D.A., Williams, A.: Facial expression of emotion and perception of the uncanny valley in virtual characters. Computers in Human behavior  \textbf{27}(2),  741--749 (2011)

\bibitem{tinwell2013perception}
Tinwell, A., Nabi, D.A., Charlton, J.P.: Perception of psychopathy and the uncanny valley in virtual characters. Computers in Human Behavior  \textbf{29}(4),  1617--1625 (2013)

\bibitem{volonte2020effects}
Volonte, M., Hsu, Y.C., Liu, K.Y., Mazer, J.P., Wong, S.K., Babu, S.V.: Effects of interacting with a crowd of emotional virtual humans on users’ affective and non-verbal behaviors. In: 2020 IEEE Conference on Virtual Reality and 3D User Interfaces (VR). pp. 293--302. IEEE (2020)

\bibitem{whitney2018ensemble}
Whitney, D., Yamanashi~Leib, A.: Ensemble perception. Annual review of psychology  \textbf{69},  105--129 (2018)

\bibitem{zell2020perception}
Zell, E., Zibrek, K., Pan, X., Gillies, M., McDonnell, R.: From perception to interaction with virtual characters. In: Eurographics (Tutorials). pp. 5--31 (2020)

\bibitem{zibrek2013evaluating}
Zibrek, K., Hoyet, L., Ruhland, K., McDonnell, R.: Evaluating the effect of emotion on gender recognition in virtual humans. In: Proceedings of the ACM Symposium on Applied Perception. pp. 45--49 (2013)

\bibitem{zibrek2015exploring}
Zibrek, K., Hoyet, L., Ruhland, K., Mcdonnell, R.: Exploring the effect of motion type and emotions on the perception of gender in virtual humans. ACM Transactions on Applied Perception (TAP)  \textbf{12}(3),  1--20 (2015)

\bibitem{zojaji2020influence}
Zojaji, S., Peters, C., Pelachaud, C.: Influence of virtual agent politeness behaviors on how users join small conversational groups. In: Proceedings of the 20th ACM International Conference on Intelligent Virtual Agents. pp.~1--8 (2020)

\end{thebibliography}

\end{document}